\documentclass[twocolumn]{aastex631}
\usepackage[utf8]{inputenc}
\usepackage{longtable}
\usepackage[flushleft]{threeparttable}
\usepackage{tabularx}
\usepackage{multirow}
\usepackage{graphicx}
\usepackage{amsmath,amssymb, fixmath}
\usepackage{color, colortbl}
\usepackage{units}
\usepackage{epstopdf}
\usepackage{hyperref}
\usepackage{multirow}
\usepackage{url}
\usepackage{subfigure}
\usepackage{rotating}
\usepackage{enumitem}\setlist[description]{font=\textendash\enskip\scshape\bfseries}
\usepackage{enumitem} \SetLabelAlign{parleft}{\parbox[t]\textwidth{#1}}%
\usepackage[multiple]{footmisc}

\usepackage{soul}
\usepackage[normalem]{ulem}

\usepackage{etoolbox}

\makeatletter
\patchcmd{\@footnotetext}{\footnotesize}{\scriptsize}{}{}
\makeatother

\newcommand{\beq}{\begin{equation}}
\newcommand{\eeq}{\end{equation}}
\newcommand{\bdm}{\begin{displaymath}}
\newcommand{\edm}{\end{displaymath}}
\newcommand{\T}[1]{\tilde{#1}}

\definecolor{Gray}{gray}{0.9}
\definecolor{orange}{rgb}{0.9,0.5,0}

\newcommand{\orcid}[1]{\href{https://orcid.org/#1}{\textcolor[HTML]{A6CE39}{\aiOrcid}}}

\begin{document}

\title{Updated Observing Scenarios and Multimessenger Implications for the International Gravitational-wave Networks O4 and O5}

\author[0000-0002-9108-5059]{R. Weizmann Kiendrebeogo}
\affiliation{Laboratoire de Physique et de Chimie de l’Environnement, Université Joseph KI-ZERBO, Ouagadougou, Burkina Faso}
\affiliation{Artemis, Observatoire de la Côte d’Azur, Université Côte d’Azur, Boulevard de l'Observatoire, F-06304 Nice, France}
\affiliation{School of Physics and Astronomy, University of Minnesota, Minneapolis, MN 55455, USA}
\author[0000-0002-6121-0285]{Amanda M. Farah}
\affiliation{Department of Physics, University of Chicago, Chicago, IL 60637, USA}
\author[0000-0003-1955-2983]{Emily M. Foley}
\affiliation{B.S. Physics, B.A. Mathematics, Wake Forest University, USA}
\affiliation{School of Physics and Astronomy, University of Minnesota, Minneapolis, MN 55455, USA}

\author[0000-0001-9763-2351]{Abigail Gray}
\affiliation{School of Physics and Astronomy, University of Minnesota, Minneapolis, MN 55455, USA}
\author[0000-0002-1275-530X]{Nina Kunert}\affiliation{Institute of Physics and Astronomy, Theoretical Astrophysics, University Potsdam, Haus 28, Karl-Liebknecht-Str. 24/25, D-14476, Potsdam, Germany}
\author[0000-0003-1357-4348]{Anna Puecher}\affiliation{Nikhef -- National Institute for Subatomic Physics, Science Park 105, 1098 XG Amsterdam, The Netherlands} \affiliation{Institute for Gravitational and Subatomic Physics (GRASP), Utrecht University, 
Princetonplein 1, 3584 CC Utrecht, The Netherlands}
\author[0009-0008-9546-2035]{Andrew Toivonen}
\affiliation{School of Physics and Astronomy, University of Minnesota, Minneapolis, MN 55455, USA}
\author[0000-0002-7826-6269]{R. Oliver VandenBerg}
\affiliation{School of Physics and Astronomy, University of Minnesota, Minneapolis, MN 55455, USA}

\author[0000-0003-3768-7515]{Shreya Anand}
\affil{Cahill Center for Astrophysics, California Institute of Technology, Pasadena CA 91125, USA}
\author[0000-0002-2184-6430]{Tom{\'a}s Ahumada}
\affil{Cahill Center for Astrophysics, California Institute of Technology, Pasadena CA 91125, USA}
\author[0000-0003-2758-159X]{Viraj Karambelkar}
\affil{Cahill Center for Astrophysics, California Institute of Technology, Pasadena CA 91125, USA}

\author[0000-0002-8262-2924]{Michael W. Coughlin}
\affiliation{School of Physics and Astronomy, University of Minnesota, Minneapolis, MN 55455, USA}
\author[0000-0003-2374-307X]{Tim Dietrich}\affiliation{Institute of Physics and Astronomy, Theoretical Astrophysics, University Potsdam, Haus 28, Karl-Liebknecht-Str. 24/25, D-14476, Potsdam, Germany}\affiliation{Max Planck Institute for Gravitational Physics (Albert Einstein Institute), Am Mühlenberg 1, Potsdam D-14476, Germany}
\author[0000-0002-8146-0177]{S. Zacharie Kam}
\affiliation{Laboratoire de Physique et de Chimie de l’Environnement, Université Joseph KI-ZERBO, Ouagadougou, Burkina Faso}
\author[0000-0001-7041-3239]{Peter T.~H.~Pang}\affiliation{Nikhef -- National Institute for Subatomic Physics, Science Park 105, 1098 XG Amsterdam, The Netherlands} \affiliation{Institute for Gravitational and Subatomic Physics (GRASP), Utrecht University, 
Princetonplein 1, 3584 CC Utrecht, The Netherlands}
\author[0000-0001-9898-5597]{Leo P. Singer}
\affiliation{Astroparticle Physics Laboratory, NASA Goddard Space Flight Center, Mail Code 661, Greenbelt, MD 20771, USA}
\author{Niharika Sravan}
\affiliation{Department of Physics, Drexel University, Philadelphia, PA 19104, USA}

\correspondingauthor{R. Weizmann Kiendrebeogo}
\email{weizmann.kiendrebeogo@oca.eu}

\begin{abstract}

 An advanced LIGO and Virgo's third observing run brought another binary neutron star merger (BNS) and the first neutron-star black hole mergers. While no confirmed kilonovae were identified in conjunction with any of these events, continued improvements of analyses surrounding GW170817 allow us to project constraints on the Hubble Constant ($H_0$), the Galactic enrichment from $r$-process nucleosynthesis, and ultra-dense matter possible from forthcoming events. Here, we describe the expected constraints based on the latest expected event rates from the international gravitational-wave network (IGWN) and analyses of GW170817. We show the expected detection rate of gravitational waves and their counterparts, as well as how sensitive potential constraints are to the observed numbers of counterparts. We intend this analysis as support for the community when creating scientifically driven electromagnetic follow-up proposals. During the next observing run O4,  we predict  an annual detection rate of electromagnetic counterparts from BNS  of  $0.43^{+0.58}_{-0.26}$  ($1.97^{+2.68}_{-1.2}$) for  the Zwicky Transient Facility  (Rubin Observatory).
\end{abstract}



\section{Introduction}

After the detection of AT2017gfo \citep{CoFo2017,SmCh2017,AbEA2017f}, associated with the binary neutron star (BNS) merger GW170817 \citep{AbEA2017b} and the short gamma-ray burst GRB170817A~\citep{GoVe2017,SaFe2017,AbEA2017e,Savchenko_2017}, there have been significant electromagnetic (EM) follow-up efforts of both further BNS detections such as GW190425 \citep{AbEA2019},  \citealt{CoAh2019b,Antier:2019pzz, Hosseinzadeh_2019, Saleem_2020, Gompertz_2020, Song:2019ddw, 2019GCN.24767....1K} and the neutron-star black hole (NSBH)  coalescences GW200105 and GW200115 \citep{AbEA2021},
\cite{AnCo2020,AnAg2020, Wang_2022, 10.1093/mnras/stac1283}.

Many science cases motivate the follow-up of neutron star mergers; these science cases include constraints on the neutron star equation of state \citep{BaJu2017, MaMe2017, CoDi2018b, CoDi2018, CoDi2019b, AnEe2018, MoWe2018,RaPe2018,Lai2019,DiCo2020,Huth:2021bsp}, the Hubble constant \citep{CoDi2019,CoAn2020,2017Natur.551...85A,2019NatAs...3..940H,Nissanke_2010,DiCo2020,PhysRevLett.122.061105,nissanke2013determining}, and $r$-process nucleosynthesis \citep{ChBe2017,2017Sci...358.1556C, CoBe2017,PiDa2017,RoFe2017,SmCh2017,WaHa2019,KaKa2019}. 
Even in the absence of further counterpart detections \citep[e.g.,][]{Andreoni2019S190510g, CoAh2019b, 2019ApJ...881L...7G, Gomez2019, LuPa2019, AnCo2020, Ackley2020, Andreoni2020S190814bv, AnAg2020,GoCu2020,KaAn2020}, these ``upper limit'' observations place constraints on possible kilonovae (KNe)  counterparts and their potential progenitor parameters \citep[e.g.,][]{CoDi2019b,KaAn2020}.

However, triggering target-of-opportunity (ToO) observations on gravitational-wave (GW) events comes at the cost of precious telescope time that could otherwise be employed for alternative science cases. Therefore, to make the most of available ToO time, we must understand how the potential targeted observations contribute to our specific scientific goals.
To support this effort, ``observing scenarios'' are produced to simulate the detection and localization of GW events, \citep[e.g.,][]{SiPr2014,AbEA2016h}.
\cite{PeSi2021} recently produced observing scenarios tuned to open public alerts from the third observing run (O3), improving the consistency of the localization performance in O3 by simulating the actual GW signal-to-noise ratio (S/N) threshold used during O3 and allowing for the inclusion of single detector searches \citep{GoEs2020,NiDe2020}. Studies like  \cite{Nissanke_2013}, \cite{PeSi2021} and \cite{Colombo:2022zzp}, which provide a set of simulated merger signals detected by the  international gravitational-wave network (IGWN)  during each observing run, lend the ability to realistically predict how well we can address specific scientific questions pertaining to the nature of compact objects, $r$-process nucleosynthesis, and the expansion rate of the Universe within the next decade.

In this paper, we describe the simulations produced for the observing scenarios currently available to the user community in the IGWN User Guide\footnote{\url{https://emfollow.docs.ligo.org/userguide/capabilities.html}}, as well as simulate potential science constraints based on self-consistent counterpart search simulations.
In Sec.~\ref{sec:scenarios}, we summarize the simulations of the observation scenarios expected during the next observation campaigns. In Sec.~\ref{sec:kilonovae}, we report the results of our simulations on the tracking of EM counterparts of GWs by optical telescopes, notably the Zwicky Transient Facility (ZTF; \citealt{Bellm:19:ZTFScheduler,Graham2018,Masci2019,DeSm2018}), a time–domain optical survey with a very wide field of view (FOV) of 47 deg$^{2}$ mounted on the Samuel Oschin 48-inch (1.2 m) Telescope at the Palomar Mountain, and the future Vera C.\ Rubin Observatory's Legacy Survey of Space and Time (Rubin Observatory; \citealt{Ivezic2019}), a large (8.4 m), wide-field (9.6 deg$^{2}$ FOV) ground-based telescope designed to conduct deep 10 yr survey of the Southern sky. In Sec.~\ref{sec:H0-discussion}, we present an estimate of how future multimessenger observations during O4 and O5 will help us to measure the Hubble constant $H_0$. We present our conclusions in Sec.~\ref{sec:conclusion}.

\section{Observing Scenarios}
\label{sec:scenarios}

\subsection{Overview}
Here, we perform detailed simulations for the upcoming fourth and fifth LIGO-Virgo-KAGRA (LIGO; \citealt{LIGO_detector}, Virgo; \citealt{Virgo_detector} and  KAGRA; \citealt{KAGRA_detector}), observing runs (O4 and O5, respectively) and present the multi-messenger constraints on the Hubble constant that can be derived from future events.
The continuous improvement of GW detector sensitivities allows probing farther into the Universe to detect more  compact binary coalescences (CBCs). 
Furthermore, since KAGRA joins the LIGO and Virgo detectors for the next observing runs, we will have a total of 4 detectors online, which might result in an increased detection probability depending on KAGRA's sensitivity. 
Following \cite{PeSi2021}, we simulate realistic \emph{astrophysical} distributions of the mass, spin, and sky locations of CBCs by assessing the 
likelihood of detection for the networks considered. We can estimate the distributions of sky-localization areas and distances that we expect for \emph{detected} events, as well as the rate of GW event detection. We will use two characteristic surveys to assess counterpart detection chances, i.e., ZTF, which will be sensitive to BNS mergers similar to GW170817/AT2017gfo up to $\sim$ 300\,Mpc \citep{CoAh2019b,AnCo2020,KaAn2020}, and Rubin Observatory  \citep{AnMa2022}, which will observe well beyond the IGWN horizon of current GW detectors. 

\begin{figure*}
\begin{center}

\includegraphics[width=7.in]{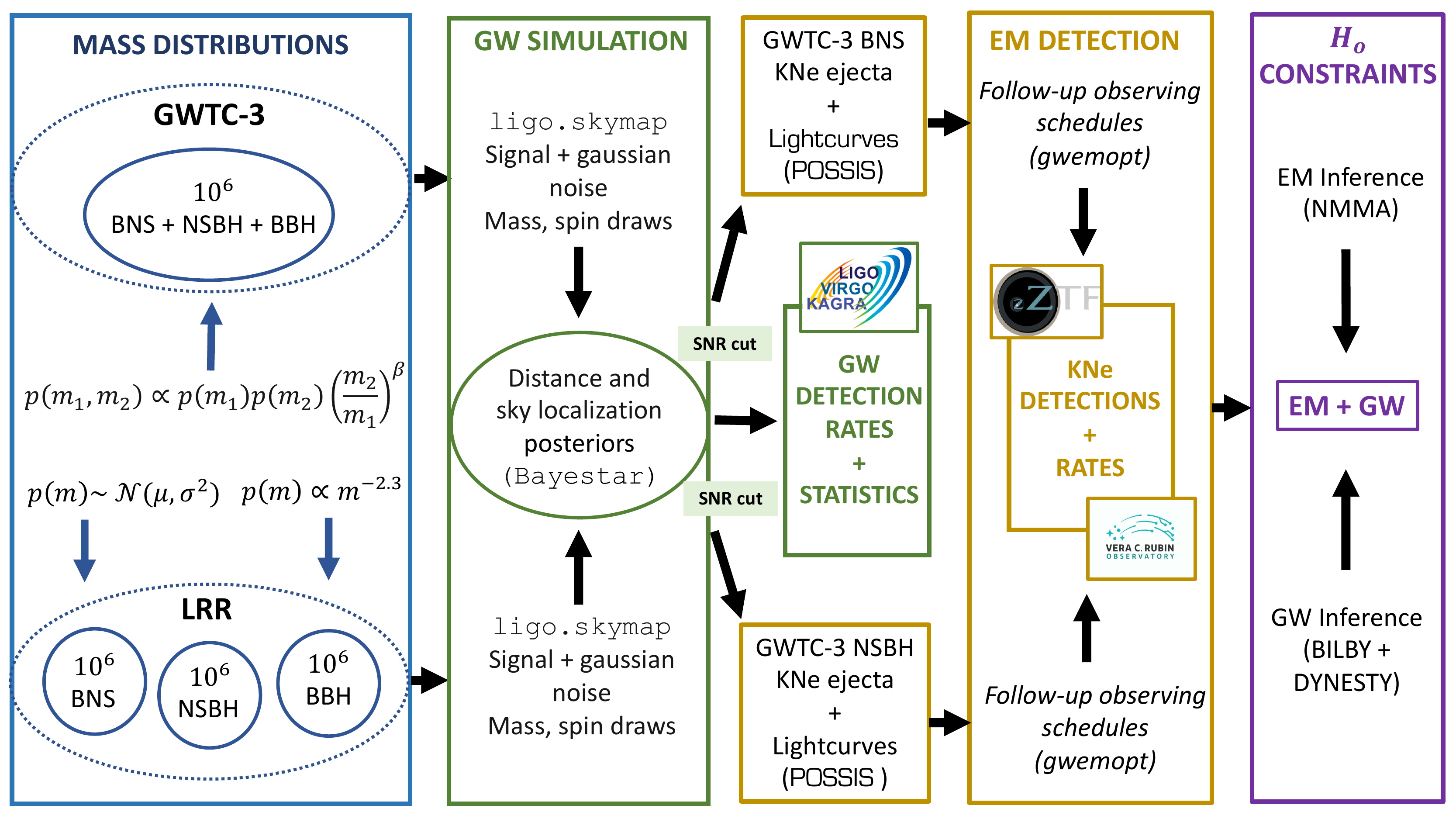}
\caption{Flowchart of observing scenarios process. Here, $\mu =1.33 {M}_\odot$, and $\sigma=0.09  {M}_\odot$, then in \texttt{LRR} case, $m$ represents the primary mass $m_1$ or secondary mass $m_2$ , since they are drawn in the same way with $m_2 \leq m_1$.}
\label{fig:flowchart}
\end{center}
\end{figure*}

\figureautorefname ~\ref{fig:flowchart} provides a flowchart for the observing scenarios pipeline and summarizes the overall process. Each step of the workflow will be further described in the following subsections.

\subsection{Population Models}
In this section, we outline the processes used to generate two distributions of CBCs which we will compare in this work. 
The first distribution, \texttt{LRR} (Living Reviews in Relativity), is drawn from the population model outlined in \cite{LRR_2020}, and \cite{PeSi2021} and used for previous IGWN observing scenarios. This distribution consists of the normal distribution for neutron star masses and the power law for black hole masses.
The second distribution, \texttt{PBD/GWTC-3} (\texttt{GWTC-3}), is derived from the population model described in \cite{Farah_2022}  and \cite{GWTC-3_2023} is used for the current observing scenarios.
Specifically, we fit the \textsc{Power Law + Dip + Break} (PDB)  model used in \cite{Farah_2022} to all CBCs in GWTC-3 from \cite{GWTC-3_2023} and then use the maximum a posteriori value of the resulting fit.

Each distribution consists of three astrophysical subpopulations of CBCs: BNS, NSBH, and BBH.
These subpopulations are separated based on the masses of their components: $m_1$ is the mass of the primary component, and $m_2$ is the mass of the secondary, with $m_2 \leq m_1$ by definition. 
The mass of a nonrotating neutron star cannot exceed the   Tolman–Oppenheimer–Volkoff (TOV)   limit ${M}_{\text{max, TOV}} \approx 2 - 2.5\,\, {M}_\odot$ \citep{PhysRevD.104.063003, PhysRevD.97.021501, PhysRevD.100.023015, 2017ApJ...850L..19M,Rezzolla:2017aly,Dietrich:2020lps}.
However, rotating NSs can exceed this limit~\citep{Baumgarte:1999cq,Stergioulas:2003yp}.
Additionally, a population analysis of all CBCs detected by the IGWN finds a sharp feature in the mass distribution of compact objects at ${2.4}_{-0.5}^{+0.5}\,\,{M}_{\odot}$ (90\% credible interval) and interprets this feature as a delineation between NSs and BHs, due to its proximity to the TOV limit~\citep{Farah_2022, GWTC-3_2023}.
In order to provide a conservative upper bound on the number of NS-containing events so that follow-up programs can make optimistic plans for observing EM counterparts, we take the 95\% upper bound on the location of the feature found by \cite{Farah_2022} and choose to delineate between NSs and BHs at $3\, {M}_\odot$.
This high value comes at the potential expense of contaminating the BNS and NSBH samples with a few low-mass BBHs, but we find this preferable to the possibility of wrongly classifying an event that could result in a bright EM counterpart as a BBH.
The choice of choosing the subpopulation boundary to be  $3\, {M}_\odot$ also maintains a consistency with previous analyses~\citep{LRR_2020}.

\begin{figure}[ht!]
\centering
\includegraphics[scale=0.42]{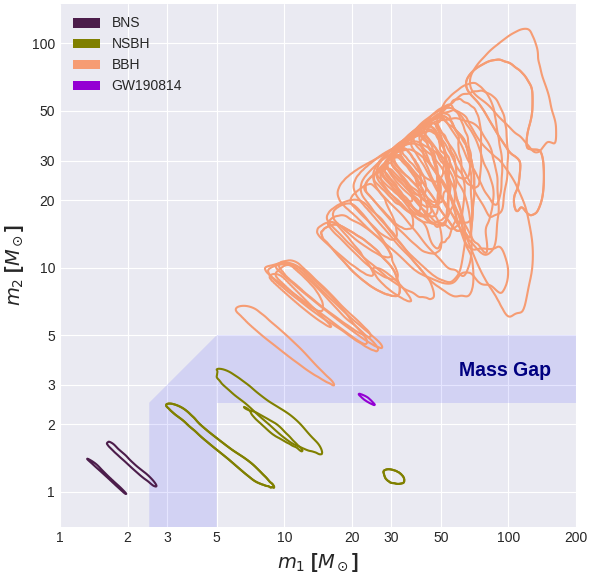}
\caption{90\% posterior credible intervals for the component masses for all CBCs in the GWTC-3 catalog \citep{GWTC-3_2023}  study assuming uniform priors in detector-frame masses and fixed FAR about 0.25 year$^{-1}$  \citep{GWTC-3_2023}. Events classified by the LIGO-Virgo-KAGRA collaboration as BNSs, NSBHs, and BBHs are shown in dark violet, olive (dark yellowish-green), and  orange, respectively. The ambiguously classified event GW190814 is shown in dark purple. The gray band indicates the approximate location of the purported \emph{lower-mass gap}. GW190814 is the only event within this region at more than 90\% credibility}
\label{fig:farah-mass-CBCs}
\end{figure}

We follow \cite{Farah_2022}, who proposed a resolution to thisNS–BH  discrepancy by using all publicly available CBCs in the GWTC-2.1 catalog in a single population analysis, thereby foregoing the need for a priori classifications and instead allowing the population fit to pick out distinct subpopulations of CBCs.  We use a similar procedure.  Figure~\ref{fig:farah-mass-CBCs} shows all publicly available CBCs in the GWTC-3 catalog \citep{GWTC-3_2023}.

\texttt{LRR} \emph{distribution}. Here, as in \cite{PeSi2021} and \cite{LRR_2020}, we use separate models to describe each CBC subpopulation (BNS, NSBH, and BBH). For  the BNS population, we draw from a truncated Gaussian mass distribution centered at $1.33\,\, {M}_\odot$ \citep{_zel_2016} from the interval $[1, 3]\,\, {M}_\odot$ with a standard deviation of $0.09\,\, {M}_\odot$ ( $p(m) \sim \mathcal{N}(\mu,\,\sigma^{2})$ ) and spin uniformly distributed magnitudes in the interval $[0, 0.05]$. For the BBHs, sampling is performed from  $[3, 50]\,\, {M}_\odot$ \citep{Abbott_2019} using a truncated power-law distribution $p(m) \propto \text{m}^{a}$ with $a =-2.3$~\citep{salpeter1955luminosity}. Both masses are independently drawn from this distribution and paired according to the rule $m_2\leq m_1$.
The spins are also uniformly distributed, with magnitudes smaller than $0.99$. In both cases, the spins are either aligned or antialigned, i.e., we neglect the possibility of misaligned spin and precessing systems in this work. Lastly, the NSBH population mass and spin distributions are described by drawing one component each from the BNS and BBH distributions above.
For each of the three subpopulations, we draw $10^6$ samples.

\texttt{GWTC-3} \emph{distribution}. This distribution is drawn from a model that describes the full population as a continuous function, foregoing the need to specify different models for each individual subpopulation~\citep{2020ApJ...899L...8F}.
The mass and spin distributions are described by the PDB model from~\cite{Farah_2022}, and \cite{GWTC-3_2023}.
This model consists of a broken power law with a notch filter $ n(m| M^{\mathrm{gap}}_{\text{low}}, M^{\mathrm{gap}}_{\text{high}}, A)$  that suppresses the merger rate between NSs and BHs ($M^{\mathrm{gap}}_{\text{low}}$ and $ M^{\mathrm{gap}}_{\text{high}}$) by a factor of $A$~\citep{2020ApJ...899L...8F,Farah_2022,GWTC-3_2023}, allowing for a potential lower-mass gap in that region~\citep{_zel_2010, 2011ApJ...741..103F}.
It additionally includes a low-pass filter at the upper end of masses of black holes to take into account a possible tapering of the mass distribution at these locations.
The component mass distribution is  then as follows:
\begin{equation}
\begin{aligned}
p(m|\lambda)  \propto &\,
 n(m| M^{\mathrm{gap}}_{\text{low}}, M^{\mathrm{gap}}_{\text{high}}, \eta_\mathrm{low}, \eta_\mathrm{high}, A) \, \times h(m|M_{\text{min}},\eta_{\text{min}}) \\
            & \times l(m|M_{\text{max}},\eta_{\text{max}})  \\
            &   \times\begin{cases}
           & \left(m/M^{\mathrm{gap}}_{\text{high}}\right)^{\alpha_1}\text{ if }m < M^{\mathrm{gap}}_{\text{high}} \\
           &\\
           & \left(m/M^{\mathrm{gap}}_{\text{high}}\right)^{\alpha_2}\text{ if }m \geq M^{\mathrm{gap}}_{\text{high}} 
          \label{Eq:farah_distribution}
       \end{cases}.
   \end{aligned}
\end{equation}
With $1 \leq m / {M}_\odot \leq 100$; \\

here, 
$n(m| M^\mathrm{gap}_\mathrm{low}, M^\mathrm{gap}_\mathrm{high}, \eta_\mathrm{low}, \eta_\mathrm{high}, A) = \Big( 1 - A h(m|M^\mathrm{gap}_\mathrm{low}, \eta_\mathrm{low})  l(m|M^\mathrm{gap}_\mathrm{high}, \eta_\mathrm{high})  \Big)$  ; \\

where
\begin{equation*}
    \begin{aligned}
        & h(m|M_{\text{min}},\eta_{\text{min}}) = 1 - l(m|M_\mathrm{min},\eta_\mathrm{min})  \\
        &\\
        & l(m|M_{\text{max}},\eta_{\text{max}})=  \left(1 + \left(m / M_\mathrm{max}\right)^{\eta_\mathrm{max}}\right)^{-1}.
   \end{aligned}
\end{equation*}

\begin{figure}[ht!]
    \centering
    \includegraphics[scale=0.5]{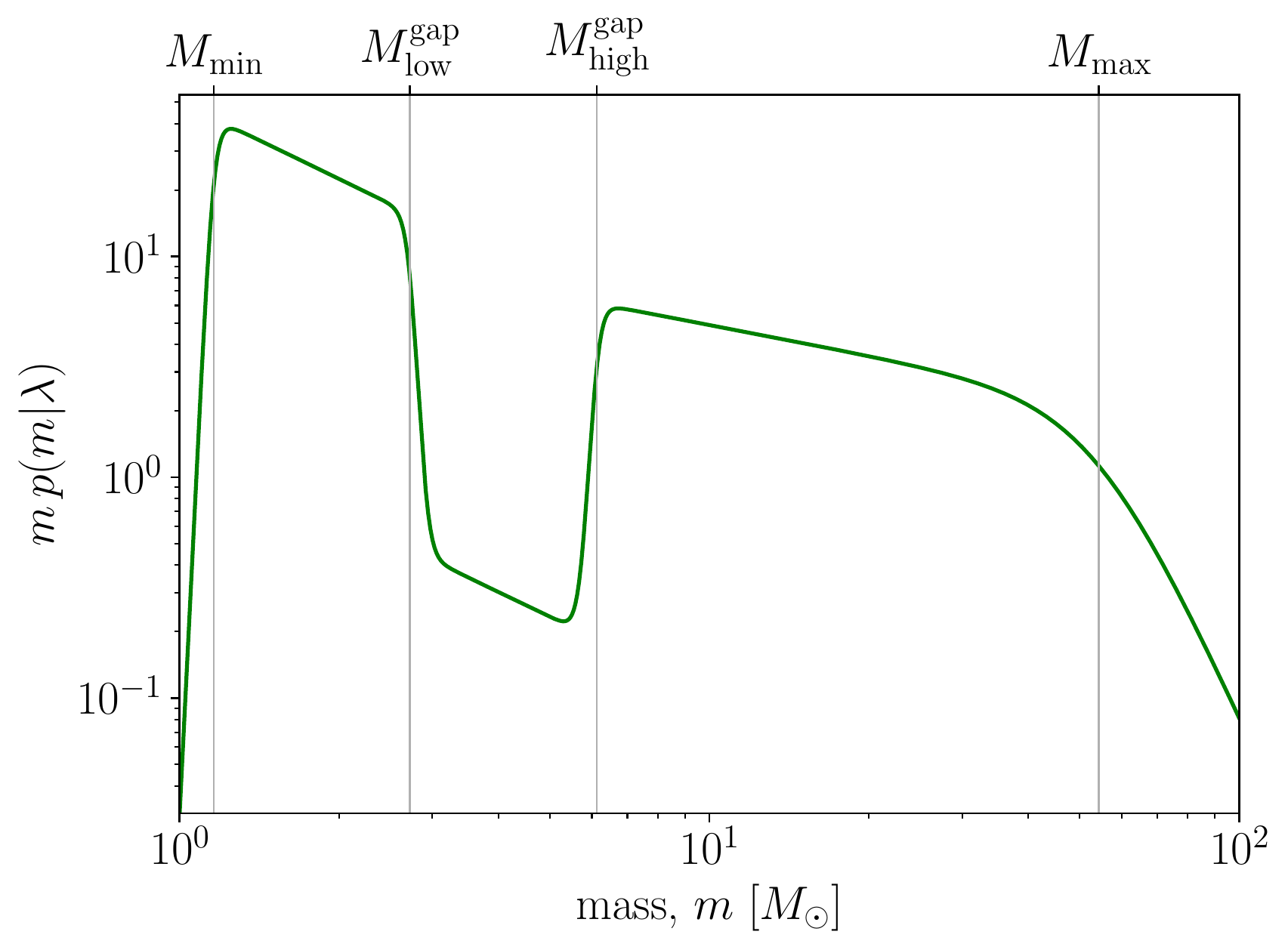}
    \caption{The 1D PDB mass distribution, $p(m|\lambda)$ on the interval $[1, 100]\,\, {M}_\odot$ for a specific choice of hyperparameters $\lambda$. See Tab.~\ref{tab:hyperpe-vals} in Appendix~\ref{App:Values_of_Hyperparam}, for the other parameters of the mass distribution.}
    \label{fig:farah_masses_distribution}
\end{figure}

$h(m|m_{\text{min}},\eta_{\text{min}})$ and $l(m|m_{\text{max}},\eta_{\text{max}})$ are the low-mass and high-mass tapering functions, respectively. \\
The 1D mass distribution, $p(m|\lambda)$, is shown in Figure~\ref{fig:farah_masses_distribution} for a specific choice of $\lambda$ ($\lambda$ represents the 12 parameters of the model; see\ Tab.~\ref{tab:hyperpe-vals} in Appendix~\ref{App:Values_of_Hyperparam}).

The 2D mass distribution is constructed by assuming that both the primary and secondary masses are drawn from $p(m|\lambda)$ and related via a "pairing function"~\citep{2020ApJ...891L..27F,Doctor2019}.
As defined in~\cite{2020ApJ...891L..27F}, the pairing assumed here is a power law in the mass ratio, $q\equiv m_1/m_2$.
Explicitly,
\begin{align}
p(m_1,m_2|\Lambda)\propto\, p(m=m_2|\lambda) p(m=m_1|\lambda) \left(\frac{m_2}{m_1}\right)^{\beta} .
   \label{eq:pairing-func}
\end{align}
The values of the hyperparameters $\Lambda = \{\lambda, \beta\}$ are listed in Appendix~\ref{App:Values_of_Hyperparam} and were chosen by fitting this model to all CBCs in GWTC-3 and choosing the maximum a posteriori value for $\Lambda$.
The effects of neglecting the hyperparameter uncertainty are estimated in Appendix~\ref{App:Values_of_Hyperparam}.

The PDB model assumes a spin distribution with isotropically oriented component spins and uniform component spin magnitudes. The spin magnitude distribution for objects with masses less than ($m < 2.5\,\, {M}_\odot$) is defined in the range of $[0, 0.4]$, and that for objects with masses larger than $2.5\,\, {M}_\odot$ is defined in the range $[0, 1]$.

\begin{figure}[ht!]
    \centering
    \includegraphics[scale=0.45]{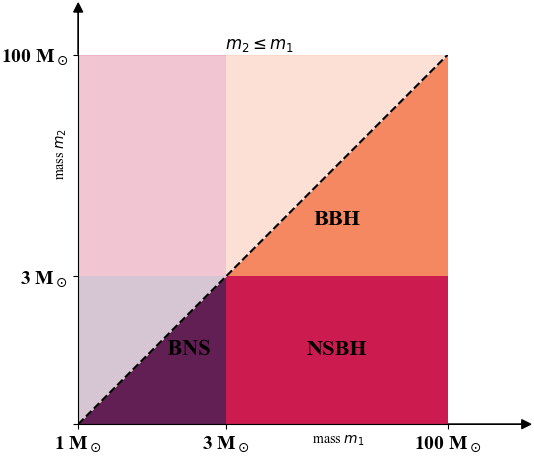}
    \caption{Cutoffs between subpopulations of compact binary coalescences for both the \texttt{GWTC-3} and \texttt{LRR} samples. It should be noted that the BHs in the \texttt{LRR} field are limited to be below $50\,\, {M}_\odot$, but are allowed to be as massive as $100\,\, {M}_\odot$ in the \texttt{GWTC-3} distribution.}
    \label{fig:masses_distribution}
\end{figure}

A set of $10^6$ CBCs were drawn from the PDB model, constituting the \texttt{GWTC-3} distribution.
These samples were then split into the three subpopulations by defining neutron stars as objects with masses below $3\,\, {M}_\odot$ and black holes as objects with masses above $3\,\, {M}_\odot$.
This yields $892,762$ BNSs, $35,962$ NSBHs, and $71,276$ BBHs. 
One resulting difference between the \texttt{LRR} and \texttt{GWTC-3} distributions is that the \texttt{LRR} distribution is drawn from a model defined only below $50\, {M}_\odot$ whereas the \texttt{GWTC-3} distribution is drawn from a model defined up to $100\, {M}_\odot$, allowing for higher-mass black holes in the latter case (though the tapering of the PDB mass distribution above $M_\mathrm{max} = 54.38$ does somewhat limit the number of high-mass black holes).

\begin{figure*}
    \centering
        \includegraphics[scale=0.5]{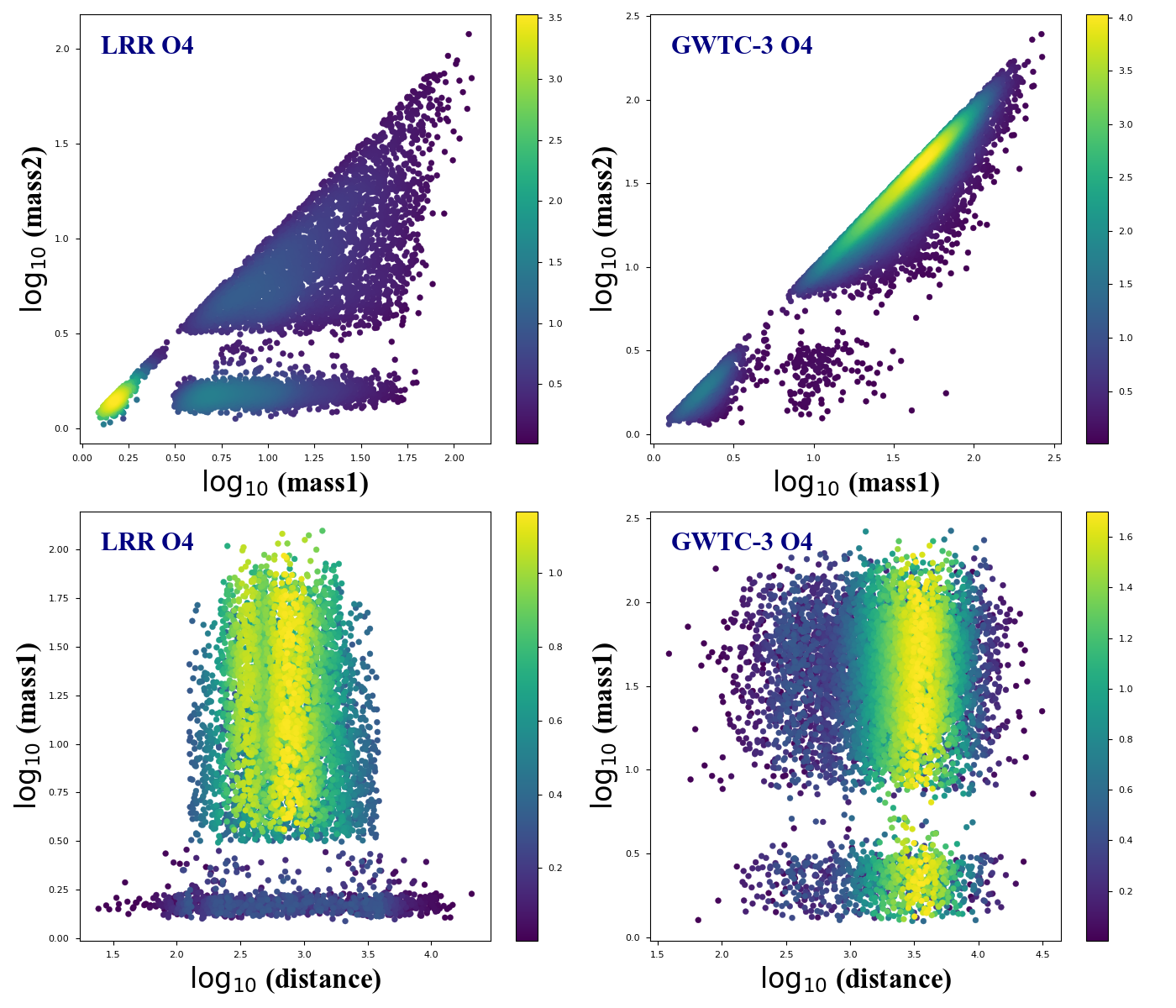}
    \caption{Simulated mass distributions for O4. The left panel shows draws from the \texttt{LRR} distribution, and the right panel shows draws from the \texttt{PBD/GWTC-3} (\texttt{GWTC-3}) distribution. The upper panels are the 2D mass distributions of the components of each CBC in the context of the detector, and the lower panels are the 2D primary mass and distance distributions. All axes are shown on a logarithmic scale. The color base shows the number of CBC events per pixel. For  O5 results, see in Appendix \ref{App:pop-dist-O5}, Figure~\ref{fig:pop-dist-O5}.
    }
    \label{fig:pop-dist-O4}
\end{figure*}
Figure~\ref{fig:masses_distribution} shows the mass distributions of the CBCs subpopulations. Figure~\ref{fig:pop-dist-O4} shows the simulated mass distributions for each model that survives the S/N cut (see Section~\ref{subsec:simulation}).

\subsection{Simulation Campaign}
\label{subsec:simulation}

We use the public software suite \texttt{ligo.skymap}\footnote{\url{https://git.ligo.org/leo-singer/ligo.skymap}}, which provides tools to read, write, generate, and visualize GW sky maps from the IGWN. After having drawn the intrinsic parameters, masses and spins of the CBCs from each of our distributions (\texttt{LRR} and \texttt{PBD/GWTC-3} (\texttt{GWTC-3})), we distribute all samples uniformly in comoving volume and isotropically in both sky location and orbital orientation. This choice reflects our expectation that GW sources are not spatially clustered or preferentially facing toward or away from Earth \citep{Ade:2015xua}.
We add Gaussian noise to the simulation of the GW signals of our CBCs. 
All the source code to reproduce these simulations, from the drawing of intrinsic parameters\footnote{\scriptsize{\url{https://lscsoft.docs.ligo.org/ligo.skymap/tool/bayestar_inject.html}}} to the statistical production of sky maps\footnote{\scriptsize{\url{https://lscsoft.docs.ligo.org/ligo.skymap/tool/ligo_skymap_stats.html}}}, passing successively by the filtering of CBC events that pass the  S/N cut \footnote{\scriptsize{\url{https://lscsoft.docs.ligo.org/ligo.skymap/tool/bayestar_localize_coincs.html}}}, as well as their location in the sky\footnote{\scriptsize{\url{https://lscsoft.docs.ligo.org/ligo.skymap/tool/bayestar_localize_coincs.html}}} are publicly accessible on GitHub\footnote{\scriptsize{\url{https://github.com/lpsinger/observing-scenarios-simulations/tree/v1}}} for \texttt{LRR} distribution, and on GitHub\footnote{\scriptsize{\url{https://github.com/lpsinger/observing-scenarios-simulations/tree/v2}}}/ \citep{leo_singer_2022_7305534} for \texttt{GWTC-3}.

Following \cite{PeSi2021}, we apply an S/N threshold of 8 for the entire \texttt{GWTC-3} distribution, and the BNSs and NSBHs populations of the \texttt{LRR} distribution. This S/N threshold is set to 9 for the BBH population of the \texttt{LRR} distribution, consistent with the localization area and distance distributions of O3 alerts \citep{PeSi2021}.
This simulation set yields estimates of the GW sky-localization area for all subpopulations, the luminosity distance, and the comoving volume. We provide a 90$\%$ credible prediction of the comoving region and volume, containing the total posterior probability. As in \cite{LRR_2020} and \cite{PeSi2021}, the localization of the sky area is provided by \texttt{Bayestar}, the rapid localization code used in production IGWN alerts \citep{SiPr2016}.

According to the IGWN, four detectors, namely LIGO Hanford, LIGO Livingston, Virgo, and KAGRA \citep{KAGRA:2013rdx}, will be engaged during the next two observing campaigns, O4 and O5. In our simulation, we adopt this configuration, along with the assumption that the four detectors each have a $70\%$ operating cycle, independently of each other. However, the recent update states that KAGRA will start  the run with LIGO Hanford, LIGO Livingston and Virgo, then return to extended commissioning to rejoin with greater sensitivity late in O4. The noise power spectral density (PSD), also known as sensitivity curves, is applied to each observation run and for each detector\footnote{\scriptsize{\url{https://observing.docs.ligo.org/plan/index.html}}}.
We use the publicly available noise curves released in LIGO-T2200043-v3\footnote{\scriptsize{\url{https://dcc.ligo.org/T2200043-v3/public}}}.
For O4, we used \texttt{aligo\_O4high.txt, avirgo\_O4high\_NEW.txt}, \texttt{kagra\_10Mpc.txt} respectively for LIGO (LHO, LLO), Virgo, and KAGRA respectively, while, for O5, we used \texttt{AplusDesign.txt}, \texttt{avirgo\_O5low\_NEW.txt}, and \texttt{kagra\_128Mpc.txt}.

In order to measure the performance of the different interferometers, we used  those sensitivities to calculate the BNS inspiral range  of 1.4 ${M}_\odot$ binary system detected with S/N = 8, during the next observation O4 and O5. The distances (in megaparsecs) from  the BNS inspiral range are recorded in Table~\ref{tab:BNS-inspiral-range}.

For the simulations, we must assume an astrophysical merger rate (taken to be in a frame  that is comoving with the Hubble flow). As in \cite{PeSi2021}, it is averaged over an assumed nonevolving mass and spin distribution. In this set, we use the  merger rate per unit comoving volume per unit proper time provided by the PDB (pair) model in the first row of Table II in \cite{GWTC-3_2023} to standardize our merger rates. PDB (pair) model uses the mass and spin distribution that is the closest match  to \texttt{GWTC-3}  distribution. 

We normalize the initial distribution of the \texttt{GWTC-3} with the total rate density of mergers, integrated across all masses and spins, taken to be fixed at $240_{-140}^{+270}\,\mathrm{Gpc}^{-3}\mathrm{yr}^{-1}$ (which is in the first row and last column on this table). For the \texttt{LRR} distributions,  we also used  PDB (pair) model, by taking the numbers from the first line and the columns (1), (2), and (3) of the same Table II  \cite{GWTC-3_2023}. We reproduce these astrophysical density rates in Table~\ref{tab:merger-rate-density}. 

\begin{table}[h!]
\renewcommand\arraystretch{1.3}
\setlength{\tabcolsep}{0.13cm}
    \centering
    \caption{The Cosmology-corrected Inspiral Sensitive Distance (in Mpc) from a GW Strain PSD}
    \begin{tabular}{cccc}
    \hline\hline
        \textbf{Run} &\ \textbf{LIGO (L1 - H1)} &\ \textbf{Virgo (V1)} &\ \textbf{KAGRA (K1)} \\ 
    \hline\hline

    \multicolumn{4}{c}{BNS Inspiral Range of the in Mpc}\\
    \hline
    
    \textbf{O4}  &$224$ &$145$ &$37$\\

\hline
\textbf{O5}     &$494$ &$183$ &$390$\\
\hline\hline                          
    \end{tabular}
    \label{tab:BNS-inspiral-range}
\end{table}

\begin{table}[ht!]
\renewcommand\arraystretch{1.4}
\setlength{\tabcolsep}{0.3cm}
    \centering
    \caption{The merger rate per unit Comoving Volume Used for \texttt{LRR} and \texttt{PBD/GWTC-3} (\texttt{GWTC-3}) Distribustions.}
    \begin{tabular}{cccc}
    \hline\hline
  \textbf{Distribution} &\ \textbf{BNS} &\ \textbf{NSBH} &\ \textbf{BBH}\\ 
    \hline\hline
    \multicolumn{4}{c}{ Merger Rate Density (Gpc$^{-3}$ yr$^{-1}$)}\\
    \hline
      \multirow{2}*{}         \texttt{LRR}     &$170^{+270}_{-120}$      &$27^{+31}_{-17}$    &$25^{+10}_{-7}$ \\
                    
                             \texttt{GWTC-3}  &$210^{+240}_{-120}$    &$8.6^{+9.7}_{-5}$ &$17.1^{+19.2}_{-10}$  \\
    \hline\hline                     
    \end{tabular}
    \label{tab:merger-rate-density}
\end{table}

\subsection{Results}
\label{subsec:observing-scenarios-results}

We make the results of the observing scenarios, including the sky-localization FITS files, available on for \texttt{PBD/GWTC-3} (\texttt{GWTC-3}) \citep{michael_w_coughlin_2022_7026209} and \texttt{LRR} \citep{r_weizmann_kiendrebeogo_2023_7623166} in separate repositories.
These simulations allow us to estimate the rates of detection by the IGWN over O4 and O5; in the following, we will focus on O4 as an example for follow-up simulations by ZTF and Rubin Observatory (although, in practice, Rubin Observatory is not expected to contribute significantly until O5). However, we reproduce some of the analyses for O5, and report them in  Appendix~\ref{App:appendix_A}. 

\subsubsection{Detection rates and summary statistics for O4 and O5}
\label{subsec:CBCs}

\begin{figure}
    \centering
    \includegraphics[scale=0.27]{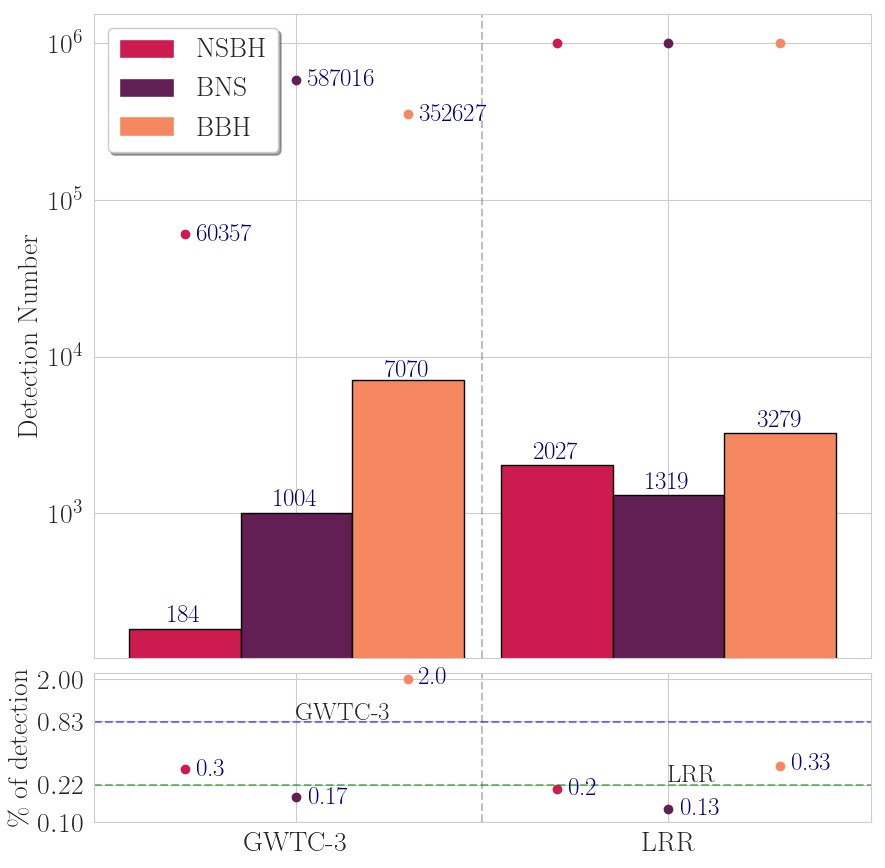}
    \caption{
         This figure shows the simulated detections for O4. The upper panel shows 
         the number of injections (CBCs from each 
         population drawn from both 
         distributions) in colored dots and the 
         bar chart representing the number of 
         events passing the S/N cut. The 
         bottom panel shows the percentage of detection relative to the number 
         injections for the two 
         distributions. 
         The colored dots represent the percentages of each population that passed the S/N,
         while the blue and green lines with respectively $0.83\%$ and $0.22\%$ 
         are successively percentages of detection of all the events (BNS + NSBH + BBH) of the 
         \texttt{PBD/GWTC-3} (\texttt{GWTC-3}) and \texttt{LRR} distributions injected in our simulation.
         For O5,  see Appendix \ref{App:detection-number-O5}, Figure~\ref{fig:detection-number-O5}}
    \label{fig:detection-number-O4}
\end{figure}

In  Figure~\ref{fig:detection-number-O4}, we summarize the detection results of the simulation set. We provide predictions for the annual detection rates of CBCs in O4 and O5 for both the \texttt{LRR} and \texttt{GWTC-3} distributions in Table~\ref{tab:annual-merger-rate}.
The confidence interval combines both the \emph{log-normal} distribution of the merger rate and uncertainties from the Poisson counting statistics.
The low number of NSBHs predicted by the PDB model is  due to the existence of a nearly empty mass gap in that model, combined with a pairing function~\citep{2020ApJ...891L..27F} that favors equal-mass binaries.
NSBHs must \emph{straddle} the mass gap, with one component on each side.
This leads to asymmetric mass ratios, which are in turn disfavored by the model fit, as most binaries in the population are consistent with being equal mass.
A version of the PDB model with a partially filled mass gap would predict more NSBH events relative to the other types of CBCs.

\begin{table}[h!]
\renewcommand\arraystretch{1.3}
    \centering
    \caption{Annual Detection Rates for Compact Binary Coalescences That we expect for the Runs O4 and O5.
                   These uncertainties do not incorporate the Monte Carlo method, but  only combine both the \emph{log-normal} distribution of the merger rate and the \emph{Poisson counting statistics}
            }
    \begin{tabular}{ccccc}
    \hline\hline
        \textbf{Run} &\ \textbf{Distribution} &\ \textbf{BNS} &\ \textbf{NSBH} &\ \textbf{BBH}\\ 
    \hline\hline

    \multicolumn{5}{c}{Annual Number of Detections}\\
    \hline
    
    \multirow{2}*{\textbf{O4}}  &\texttt{LRR}     &$17^{+35}_{-13}$        &$10^{+18}_{-8}$     &$46^{+23}_{-17}$\\

                                &\texttt{GWTC-3}      &$36^{+49}_{-22}$        &$6^{+11}_{-5}$      &$260^{+330}_{-150}$\\ 

\hline
 \multirow{2}*{\textbf{O5}}     &\texttt{LRR}     &$86^{+171}_{-59}$       &$48^{+71}_{-30}$    &$190^{+80}_{-58}$\\
  
                                &\texttt{GWTC-3}      &$180^{+220}_{-100}$     &$31^{+42}_{-20}$   &$870^{+1100}_{-480}$\\ 
\hline\hline      

    \end{tabular}

    \label{tab:annual-merger-rate}
\end{table}

In Table~\ref{tab:summary-stat}, we also provide statistics regarding the GW signal sky-localization area, luminosity distance, and comoving volume. 
Sky-localization area (volume) is given as the $90\%$ credible region, defined as the smallest area (volume) enclosing $90\%$ of the total posterior probability. This corresponds to the area (volume) of the sky that must be covered to have a $90\%$ chance of including the source. We have adopted the same statistical treatment process as the one used in \cite{PeSi2021}.
The results from the simulation of the \texttt{GWTC-3} distribution are also available in the IGWN Public Alerts User Guide (see footnote 1).

\begin{table}[t]
\renewcommand\arraystretch{1.35}
    \centering
    \caption{Summary Statistics for O4 and O5. \\
    These recorded values are given as 90\% credible interval calculated with the 5\% and 95\% quantile. Those uncertainties have been  described by Monte Carlo sampling.}
    \begin{tabular}{ccccc}
    \hline\hline
       \textbf{Run} &\ \textbf{Dist.} &\ \textbf{BNS} &\ \textbf{NSBH} &\ \textbf{BBH}\\ 
    \hline\hline
    \multicolumn{5}{c}{ Median $90\%$  Credible Area (deg$^{2}$)}\\
    \hline
    \multirow{2}*{\textbf{O4}}      & \texttt{LRR}    &$2100^{+150}_{-220}$    &$2090^{+130}_{-130}$   &$653^{+53}_{-36}$\\
                                    &\texttt{GWTC-3}      &$1860^{+250}_{-170}$    &$2140^{+480}_{-530}$   &$1428^{+60}_{-55}$\\

    \hline
    \multirow{2}*{\textbf{O5}}      &\texttt{LRR}     &$2050^{+100}_{-160}$    &$2110^{+100}_{-100}$    &$682^{+25}_{-30}$\\
                                    &\texttt{GWTC-3}      &$2050^{+120}_{-120}$    &$2000^{+350}_{-220}$   &$1256^{+48}_{-53}$\\
    \hline
    \multicolumn{5}{c}{Median $90\%$ Credible Comoving Volume ($10^{6}\,\, \text{Mpc}^{3}$)}\\
    \hline
    \multirow{2}*{\textbf{O4}}      &\texttt{LRR}     &$46.5^{+6.6}_{-7.0}$    &$159^{+26}_{-16}$      &$207^{+21}_{-20}$\\ 
                                    &\texttt{GWTC-3}      &$67.9^{+11.3}_{-9.9}$    &$232^{+101}_{-50}$    &$3400^{+310}_{-240}$\\
    \hline
    \multirow{2}*{\textbf{O5}}      &\texttt{LRR}     &$240^{+29}_{-26}$       &$785^{+68}_{-62}$      &$857^{+63}_{-60}$\\
                                    &\texttt{GWTC-3}      &$376^{+36}_{-40}$       &$1350^{+290}_{-300}$   &$8580^{+600}_{-550}$\\ 
    \hline
    \multicolumn{5}{c}{ Median Luminosity Distance (Mpc)}\\
    \hline
    \multirow{2}*{\textbf{O4}}      &\texttt{LRR}     &$349^{+12}_{-14}$  &$564^{+15}_{-13}$     &$1102^{+33}_{-32}$\\
                                    &\texttt{GWTC-3}      &$398^{+15}_{-14}$       &$770^{+67}_{-70}$     &$2685^{+53}_{-40}$\\
    \hline
    \multirow{2}*{\textbf{O5}}      &\texttt{LRR}     &$619^{+15}_{-19}$       &$1007^{+20}_{-22}$    &$1948^{+34}_{-24}$\\ 
                                    &\texttt{GWTC-3}      &$738^{+30}_{-25}$       &$1318^{+71}_{-100}$   &$4607^{+77}_{-82}$\\  
    \hline
    
    \multicolumn{5}{c}{ Sensitive Volume : Detection Rate/ Merger Rate: (Gpc$^3$) } \\
                                
    \hline
    \multirow{2}*{\textbf{O4}}      &\texttt{LRR}     &$0.1011^{+0.0066}_{-0.0064}$    &$0.403^{+0.021}_{-0.020}$        &$1.861^{+0.077}_{-0.074}$\\ 
                                    &\texttt{GWTC-3}      &$0.172^{+0.013}_{-0.012}$       &$0.78^{+0.14}_{-0.13}$      &$15.15^{+0.42}_{-0.41}$\\
    \hline
    \multirow{2}*{\textbf{O5}}      &\texttt{LRR}     &$0.507^{+0.027}_{-0.026}$       &$1.809^{+0.070}_{-0.068}$   &$7.62^{+0.19}_{-0.19}$\\          
                                    &\texttt{GWTC-3}      &$0.827^{+0.044}_{-0.042}$        &$3.65^{+0.47}_{-0.43}$    &$50.7 ^{+1.2}_{-1.2}$\\
    \hline
    \hline                          
    \end{tabular}
    \label{tab:summary-stat}
    \end{table}

There are notable differences that arise due to the improved mass distributions measured in \texttt{GWTC-3}, which were derived from the maximum a posteriori fit to all compact binaries detected so far;  this is vastly different from the normal distribution (NS) and power law (BH) for masses assumed by \texttt{LRR}.
This difference, coupled with the inclusion of single-detector triggers, yields large differences from previous reports.
This \texttt{GWTC-3} distribution accounts for an increase in the predicted number of detected events by a factor $\sim (0.83\% / 0.22\%) = 3.772$ for O4, then $\sim (1.22\% / 0.48\%) =2.542$ for O5.
Breaking down by population type, the estimated annual detection rate is a factor of  $\sim$2 (5) times higher in BNS (BBH) subpopulations for \texttt{GWTC-3} but drops to $\sim$ 0.6  for NSBH.
The median luminosity distance predicted by \texttt{GWTC-3} is about $\sim$ 1.14 (1.19), 1.36 (1.31), and 2.44 (2.36)  larger than the value for \texttt{LRR} for BNS, NSBH, and BBH events respectively  during O4 (O5), with the median of sensitive volume similarly increasing by $\sim$ 1.7 (1.6), 1.9 (2), and 8.14 (6.65).
For sky-localizations, the results are broadly similar to previous results. For example, during O4, \texttt{GWTC-3} predicts $\sim$ 11$\%$ smaller sky-localization area for BNS subpopulation than that from \texttt{LRR}, while in the BBH case \texttt{GWTC-3} predicts about  $\sim$ 52\% larger than the value for \texttt{LRR}.

\section{Predictions for detection rates and science with gravitational-wave counterparts}
\label{sec:kilonovae}

In this section, we use the results from Sec.~\ref{sec:scenarios} to make predictions for potential counterpart detections during O4 and O5.

\subsection{Light Curves}

\begin{figure*}[ht!]
\centering
    \includegraphics[width=3.5in,height=4in]{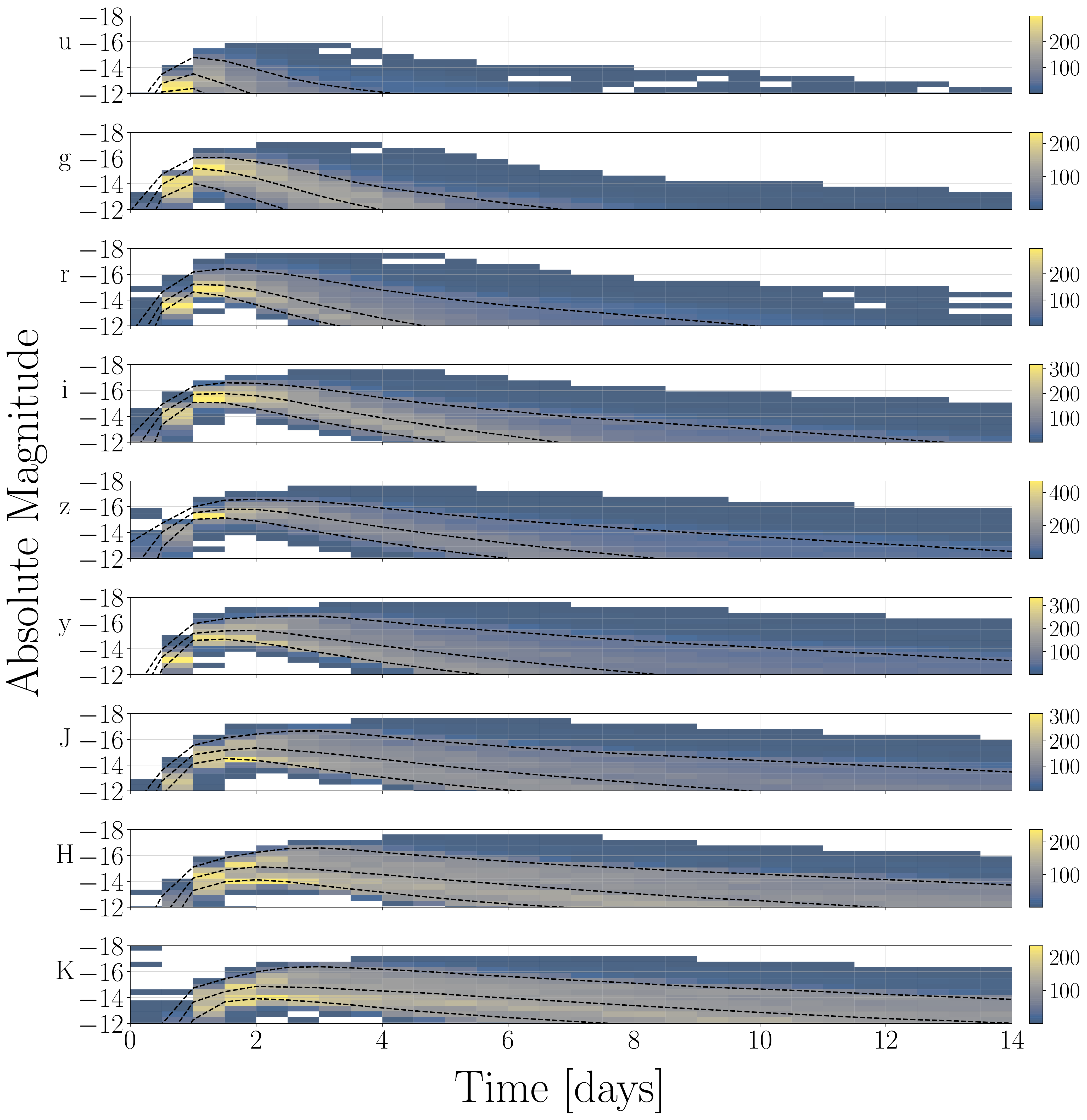}
    \hspace{0.06cm}
    \includegraphics[width=3.5in, height=4in]{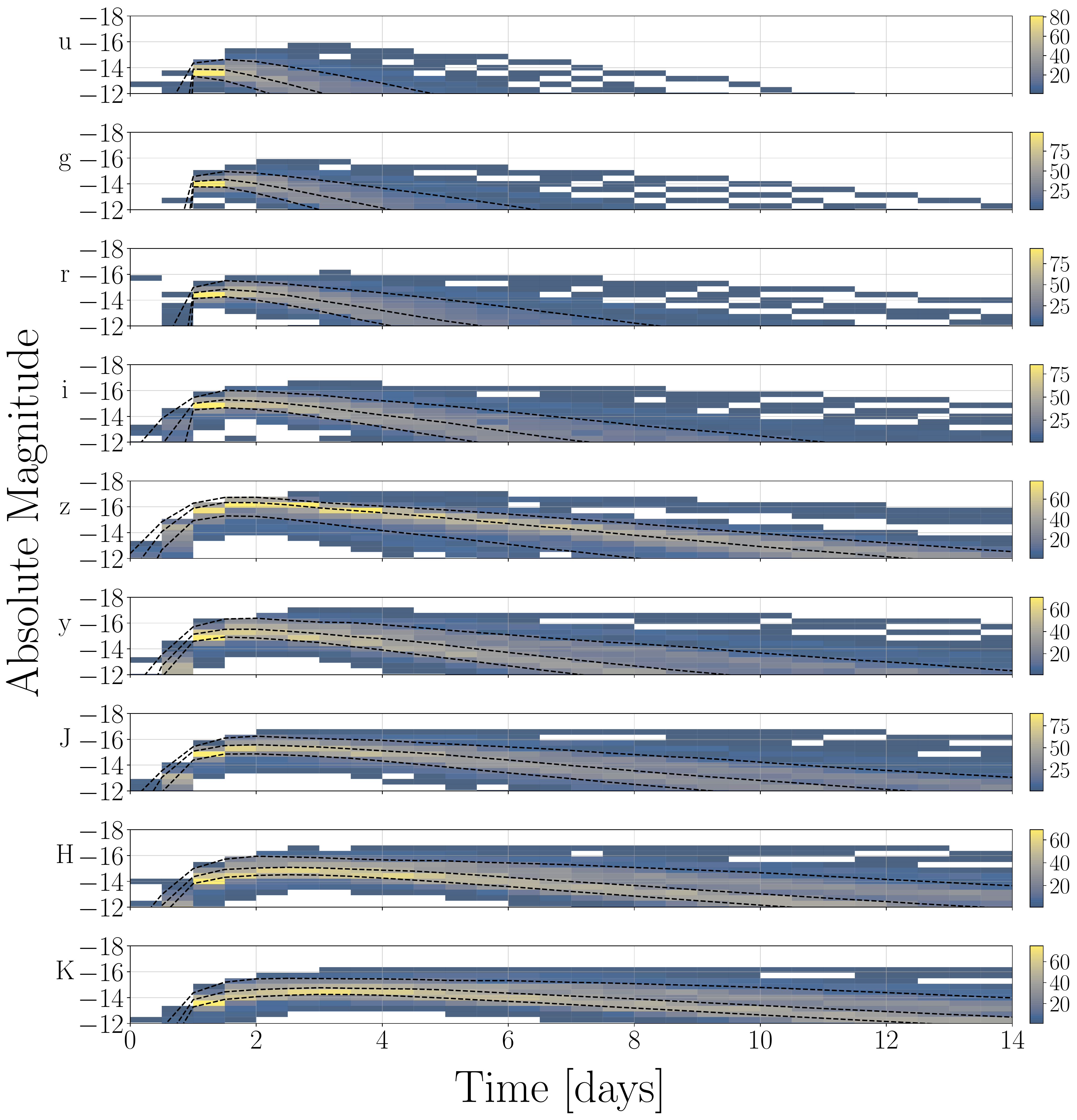}
    \caption{2-D histograms of simulated BNS (left) and NSBH (right) light curves for O4. We show light curves in $u$-, $g$-, $r$-, $i$-, $z$-, $y$-, $J$-, $H$-bands
            in order to include all the bands used by surveys considered in this work (ZTF, Rubin and Gemini). In each panel, we plot three dash lines; on top the 10th percentile, at middle the 50th percentile and on bottom the 90th percentile. The color bar shows the number of detections in the different bands.  For the observing run O5, see Figure.~\ref{fig:lightcurves-O5}, in Appendix~\ref{App:lightcurves-histogram-O5}.}
    \label{fig:lightcurves-O4}
\end{figure*}

To simulate the light curves corresponding to the BNSs and NSBHs from Sec.~\ref{sec:scenarios}, we first predict the dynamical and disk wind ejecta produced for each simulated object. For the BNS case, we use the fit from \cite{CoDi2018} for the dynamical and \cite{DiCo2020} for the disk wind. For the NSBH case, we use the fit from \cite{FoHi2018} for the dynamical and \cite{KrFo2020} for the disk wind. We note that the NSBH case often results in no KNe due to large mass ratios resulting in a direct plunge of the neutron star. To translate these ejecta quantities into light curves, we use the multidimensional Monte Carlo radiative transfer code \texttt{POSSIS} \citep{Bulla:2019muo, Bulla:2022mwo}. For the BNS case, we use the same geometry and lanthanide fractions for each component as presented in \cite{DiCo2020}, shown to yield good fits to GW170817. For the NSBH case, we use the same configurations as presented in \cite{AnCo2020}. This model takes as input the dynamical ejecta $ M^{\rm dyn}_{\rm ej}$, disk wind ejecta $M_{\rm ej}^{\mathrm{wind}}$, half opening angle $\Phi$, and the observation angle $\Theta_{\rm{obs}}$ (see Ref.~\cite{DiCo2020} for details). We take the inclination measurement from the simulated value in the observing scenarios, and we draw $\Phi$ uniformly between 15$^\circ$ and 75$^\circ$. To interpolate the grid-based model, we use a Gaussian process regression method \citep{CoDi2018,CoDi2018b}. Figure~\ref{fig:lightcurves-O4} shows a histogram of the light curves in the optical to near-infrared bands for the O4 simulation set, simulated with \texttt{NMMA}, publicly available on GitHub\footnote{\scriptsize{\url{https://github.com/nuclear-multimessenger-astronomy/nmma}}}.

\subsection{Simulated follow-up}

\begin{figure*}[ht!]
\centering
    \includegraphics[scale=0.8]{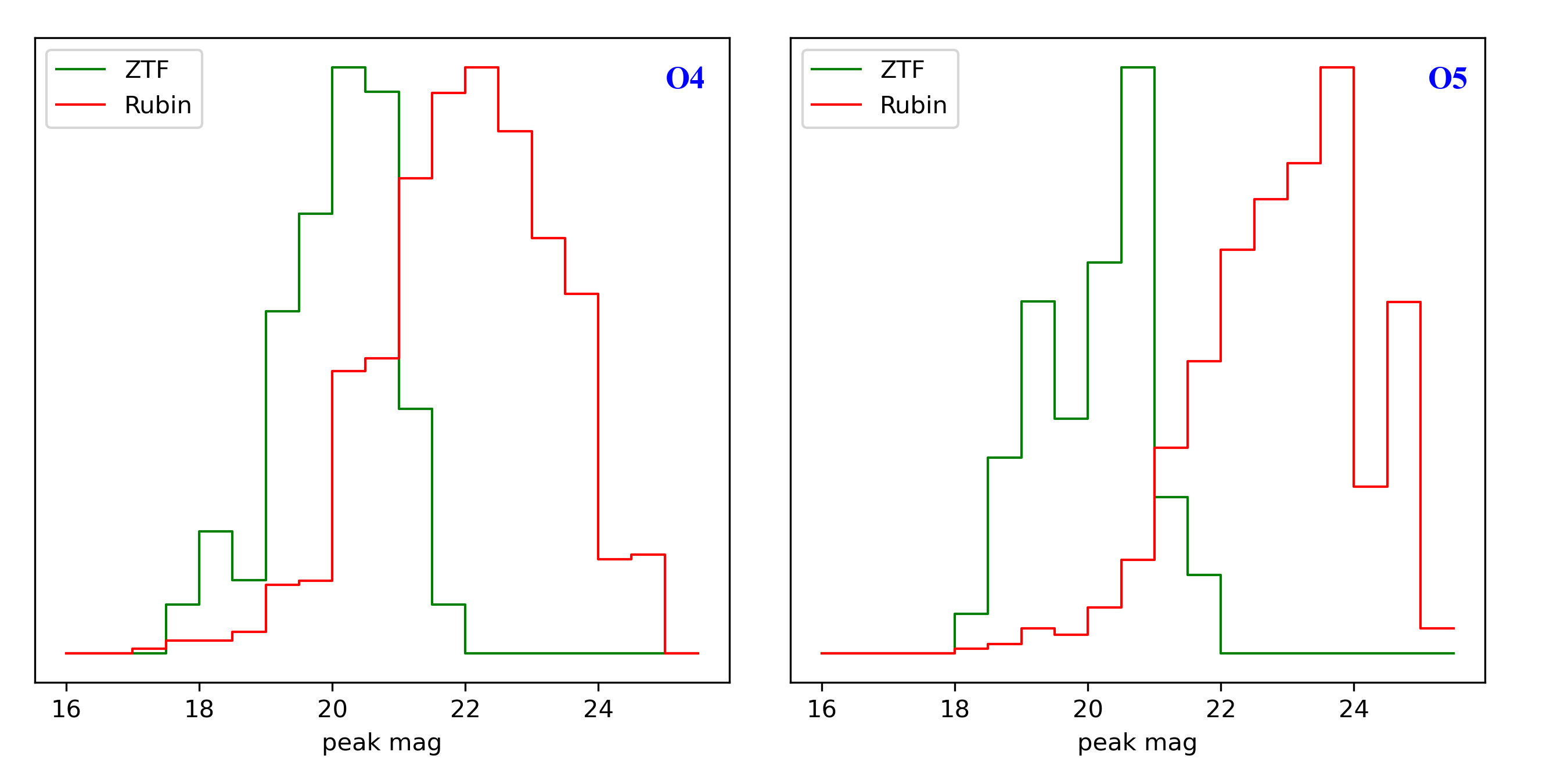}
    \caption{1D histograms of the peak magnitudes in the ZTF bands (in green) and the Rubin observatory bands (in red). On the left, we show the BNS peak mag in O4; on the right, the same plot for the run O5. See Figure~\ref{fig:peak_mag_in_each_band}, in Appendix~\ref{App:peak-mag} for the peaks related to each band of each telescope.}
    \label{fig:peak_mag_all_bands}
\end{figure*}

\begin{figure*}
    \includegraphics[width=3.5in]{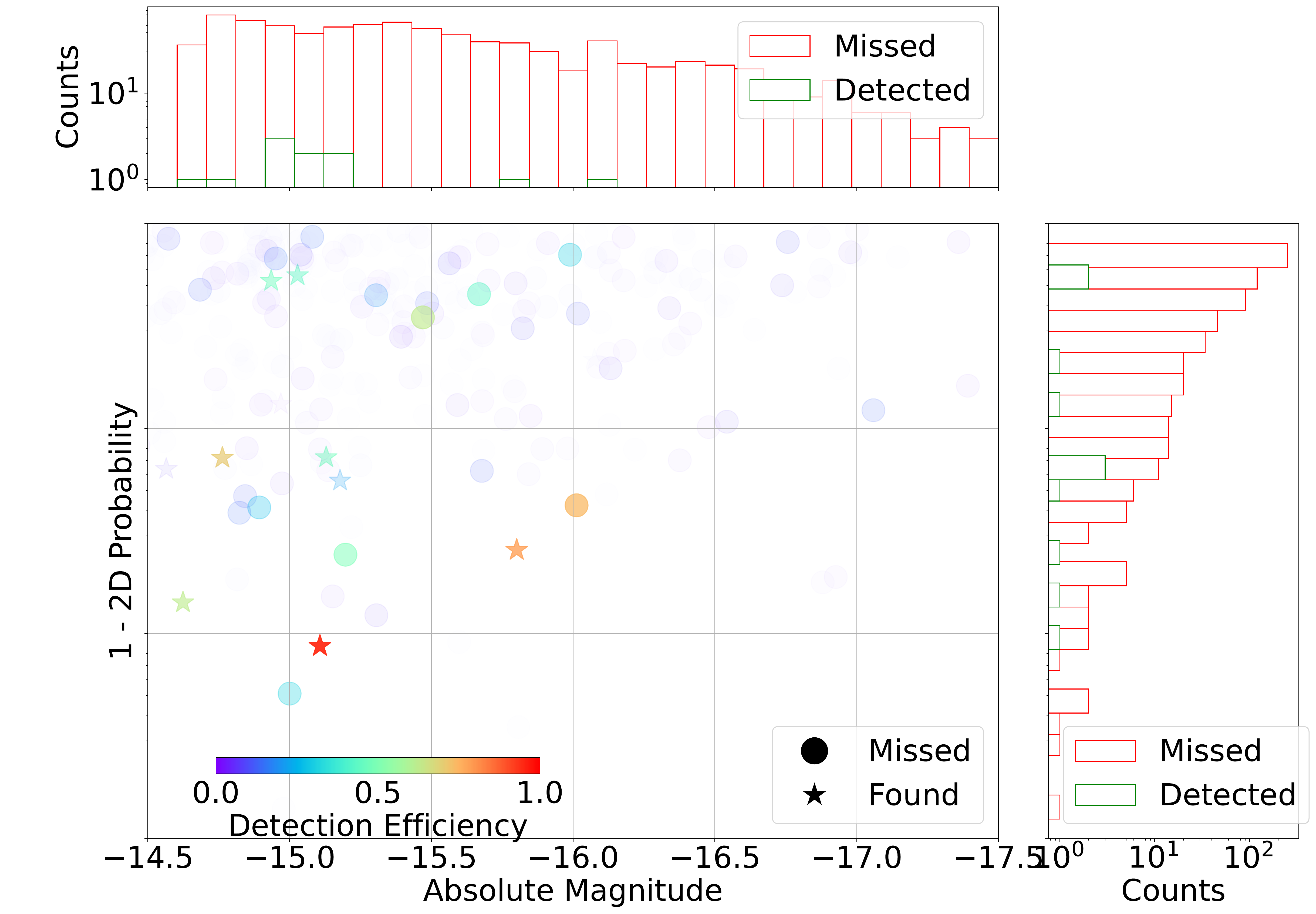}
    \includegraphics[width=3.5in]{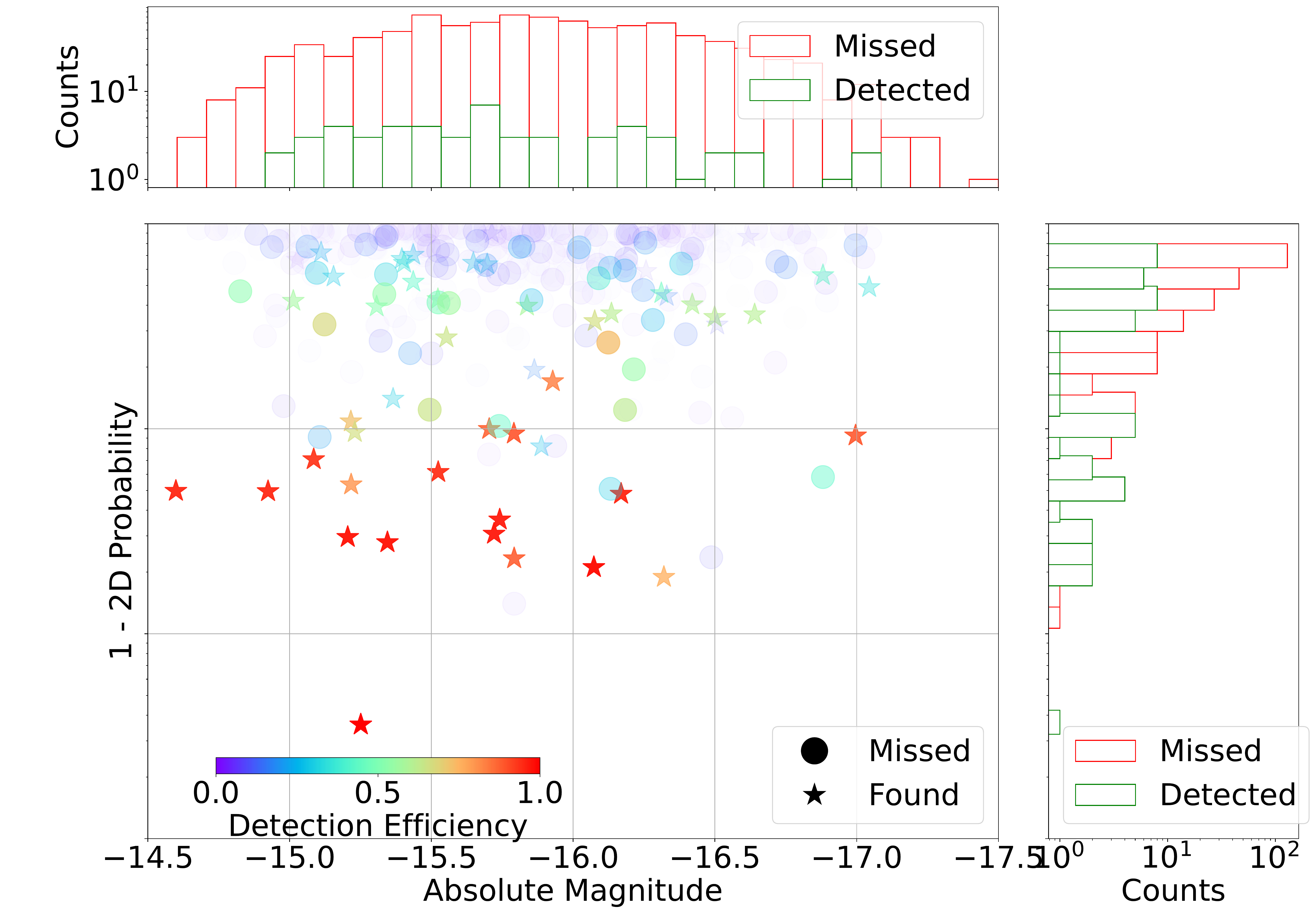}
    \includegraphics[width=3.5in]{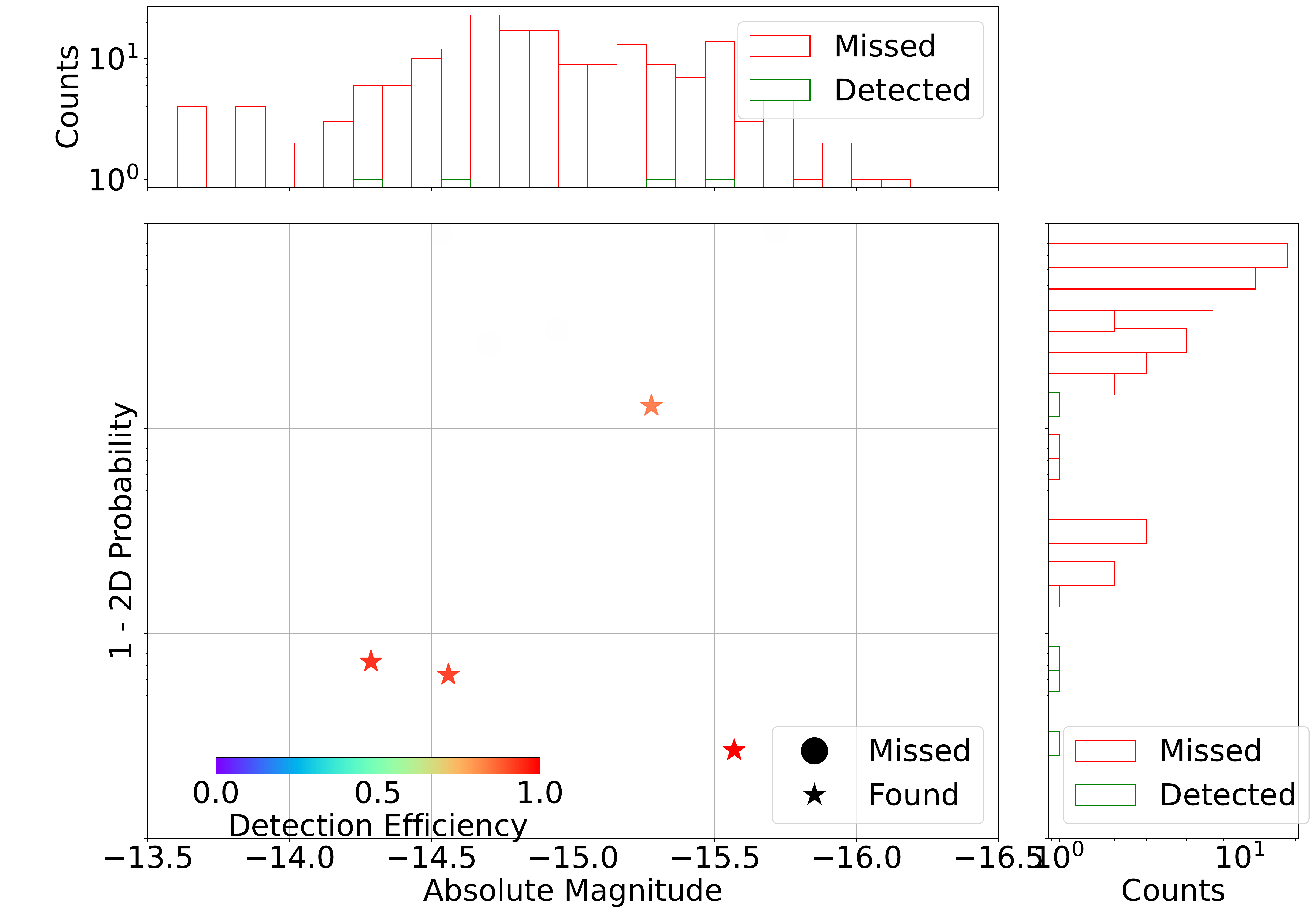}
    \includegraphics[width=3.5in]{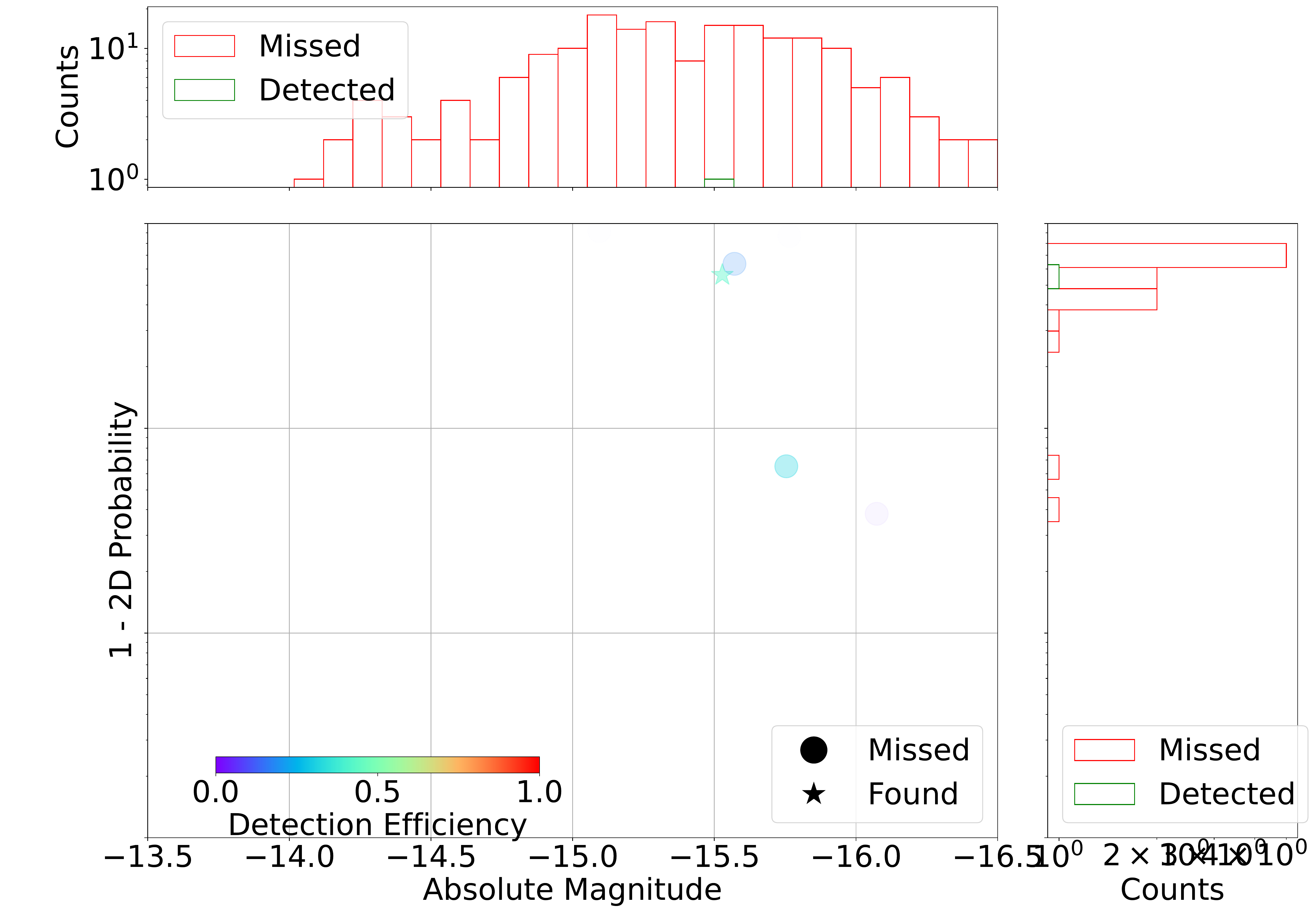}
    \caption{The plots at the top are BNS distributions of the missed and found injections (on the left, BNS light curves observed by ZTF, and the right one shows the same using Rubin Observatory), as a function of the peak $r$ band absolute magnitude and 2D probability enclosed. We encode in the color bar and the transparency of the points the 3D probability of finding the transient as measured by simulated injections. The 1D histograms are marginalized versions of the 2D scatter plot, with the detected set in green and missed set in red. The plots at the bottom are the same for the NSBH light curves. Those plots concern the run O4. In Appendix~\ref{App:detection} the Figure~\ref{fig:detection-number-O5} shows the same distribution for the run O5.}

    \label{fig:detection-O4}
\end{figure*}

We use these light curves to perform a simulation campaign for optical observations of GW counterparts with Rubin Observatory  and ZTF.
To simulate ZTF and Rubin Observatory observations, we use the Gravitational-Wave ElectroMagnetic OPTimization\footnote{\scriptsize{\url{https://github.com/mcoughlin/gwemopt}}} (\emph{gwemopt}; \citealt{CoTo2018,CoAn2019}). 
We take ZTF to have a base sensitivity in $g$, $r$, and $i$ bands with $g\sim 21.7$\,mag, $r\sim 21.4$\,mag, $i\sim 20.9$\,mag \citep{AlCo2020arXiv}, and Rubin Observatory with $u$, $g$, $r$, $i$, $z$, and $y$ bands with  $u=23.9$\,mag, $g=25.0$\,mag, $r=24.7$\,mag, $i=24.0$\,mag, $z=23.3$\,mag, and $y=22.1$\,mag. Then, in order to obtain the full range of bands used in Figure~\ref{fig:lightcurves-O4}, we incorporate Gemini Observatory, which uses sensitivity tuned $u$, $g$, $r$, $i$, $z$, $y$, $J$, $H$, and $K$ bands with $u=24.1$\,mag, $g=25.0$\,mag, $r=25.0$\,mag, $i=25.3$\,mag, $z=24.5$\,mag, $y=23.0$\,mag, $J=23.2$\,mag, $H=22.6$\,mag, and $K=22.6$\,mag.  Figure~\ref{fig:peak_mag_all_bands} shows a one dimensional histogram of prediction of the magnitude peak in all ZTF and Rubin observatory bands during the forthcoming observing runs O4 and O5.\\
\textit{gwemopt} is the package that was used to create follow-up schedules for a number of facilities during O3, including ZTF and the Dark Energy Camera, and therefore serves as reasonably realistic software to use for this purpose for O4 and O5.
We inject simulated KNe consistent with the GW localization and simulate follow-up observations, taking into account the sensitivity and FOV of the telescopes; this yields the expected fraction of detections for KNe within the simulation set.
Here, a \emph{detection} means at least one photometry in the light curve.

In our search algorithms, we sample the light curves according to ZTF Phase-II public cadence and private $i$ band survey cadence by drawing revisit cadences from kernel density estimate (KDE) fits to the same. For every visit, we assign a random night's observing filter sequence. We also add  300\,s ToO observations in $g$ and $r$ bands during the first day or first two days after the trigger, for GW localizations of greater or lesser than 1000 deg$^{2}$, respectively, by following the observations taken by ZTF during O3 \citep{KaAn2020}. These observations are executed as part of the GW EM search program within the ZTF collaboration \citep{2021NatAs...5...46A}. Given mean magnitudes, we simulate magnitude uncertainties using a skew normal fit to forced photometry uncertainties. Finally, the forced photometry upper limits are estimated using KDE fits and used to reject measurements that are fainter. We require that the light curves meet the trigger criterion of S/N $>3$ \citep{2021ApJ...918...63A}. For Rubin observatory, we used ToO observations based on the strategy presented in \cite{AnMa2022}.

For this analysis, we inject the 1004 (2003) BNS and 184 (356) NSBH of the \texttt{PBD/GWTC-3} (\texttt{GWTC-3}) distribution that have passed the S/N cutoff in O4 (O5). Figure~\ref{fig:detection-O4} shows the distribution of the missed and found events for both the BNS and NSBH cases for observing run O4.
We compare these to the peak absolute magnitude in $r$ band and the 2D probability enclosed by the simulated observations for ZTF and Rubin Observatory. There are many more BNS than NSBH detections due to the much smaller subset of NSBH injections with nonzero ejecta masses.
For comparison, we show marginalized 1D histograms for both the detected and missed sets of objects, which show distinct differences in both. The detected set shows a distinct preference toward brighter objects. It also shows a preference for both higher 2D and 3D probability coverage. 
Despite ZTF's wide FOV of 47 deg$^{2}$, its relatively lower sensitivity limits it to $\sim$ 12 (4) and  4 (1) detections of the injections respectively for BNS and NSBH events during the observing run O4 (O5). Rubin Observatory instead finds $\sim$ 55 (60) and 1 (5) of the injections for BNS and NSBH events, respectively during the next observing run O4 (O5).
By combining the EM detection fraction and the \texttt{GWTC-3} CBCs annual detection rates (Tab.~\ref{tab:annual-merger-rate} ), in Table.~\ref{tab:em-annual-detection-rate}, we provide the predictions we expect for the detection rates of EM counterparts during the forthcoming run under the specific assumptions described above.

\begin{table}[h!]
\renewcommand\arraystretch{1.3}
\setlength{\tabcolsep}{0.45cm}
    \centering
    \caption{ Annual detection rate of EM counterparts  that we expect for ZTF and Rubin Observatory during  the Run O4 and O5.}
    \begin{tabular}{cccc}
    \hline\hline
        \textbf{Run} &\ \textbf{Telescope} &\ \textbf{BNS} &\ \textbf{NSBH} \\ 
    \hline\hline

    \multicolumn{4}{c}{EM annual number of detections }\\
    \hline
    
    \multirow{2}*{\textbf{O4}}  &\texttt{ZTF}     &$0.43^{+0.58}_{-0.26}$    &$0.13^{+0.24}_{-0.11}$\\

                                &\texttt{Rubin}     &$1.97^{+2.68}_{-1.2}$    &$0.03^{+0.06}_{-0.03}$\\ 

\hline
 \multirow{2}*{\textbf{O5}}     &\texttt{ZTF}     &$0.43^{+0.44}_{-0.2}$    &$0.09^{+0.12}_{-0.06}$\\
  
                                &\texttt{Rubin}   &$5.39^{+6.59}_{-2.99}$    &$0.43^{+0.59}_{-0.28}$\\ 
\hline\hline                          
    \end{tabular}
    \label{tab:em-annual-detection-rate}
\end{table}

\subsection{KNe sample constraints}

For the objects \emph{detected} by this process, we perform parameter estimation of the resulting EM light curves.
To do so, we use the \texttt{NMMA} framework~\citep{pang2022nmma}, designed to perform Bayesian inference of multimessenger signals, incorporating all available data on the neutron star EoS in the process \citep{DiCo2020,PaTe2021,TePa2021}. 
This is an efficient package to evaluate Bayes' theorem in order to obtain posterior probability distributions, $p(\vec{\theta} | x, \mathbf{M})$, for model source parameters $\vec{\theta}$ under the hypothesis or model $\mathbf{M}$ with mock-up data $x$ as
\begin{equation}\label{Eq:Bayes_theorem}
\begin{aligned}
    p(\vec{\theta} | x, \mathbf{M}) = \frac{p(x|\vec{\theta}, \mathbf{M})\  p(\vec{\theta} | \mathbf{M})}{p(x|\mathbf{M})}\,,
\end{aligned}
\end{equation}
where $p(x|\vec{\theta}, \mathbf{M})$, $p(\vec{\theta} | \mathbf{M})$, and $p(x|\mathbf{M})$ are the likelihood, prior, and evidence, respectively.
This framework has been used in the measurement of the NS EoS and $H_0$ using GW170817 \citep{DiCo2020} and the detection of the shortest long $\gamma$-ray burst ever confirmed \citep{AhSi2021}.

\subsubsection{Ejecta constraints}


We begin by evaluating   90\% upper limits possible from the sample considered here (KNe for these objects), then constrain them for the ejecta model parameters $ M^{\rm dyn}_{\rm ej}$ and $M_{\rm ej}^{\mathrm{wind}}$, as well as a systematic contribution to the dynamical ejecta $\alpha$ and the fraction of the disk mass contributing $\zeta$. This enables us to make an empirical constraint of the fraction of the disk contributing to KNe. We constrain $ M^{\rm dyn}_{\rm ej}$  to 10--40\% and $M_{\rm ej}^{\mathrm{wind}}$ to 10--30\%.

\begin{figure}[ht!]

    \includegraphics[width=3.5in]{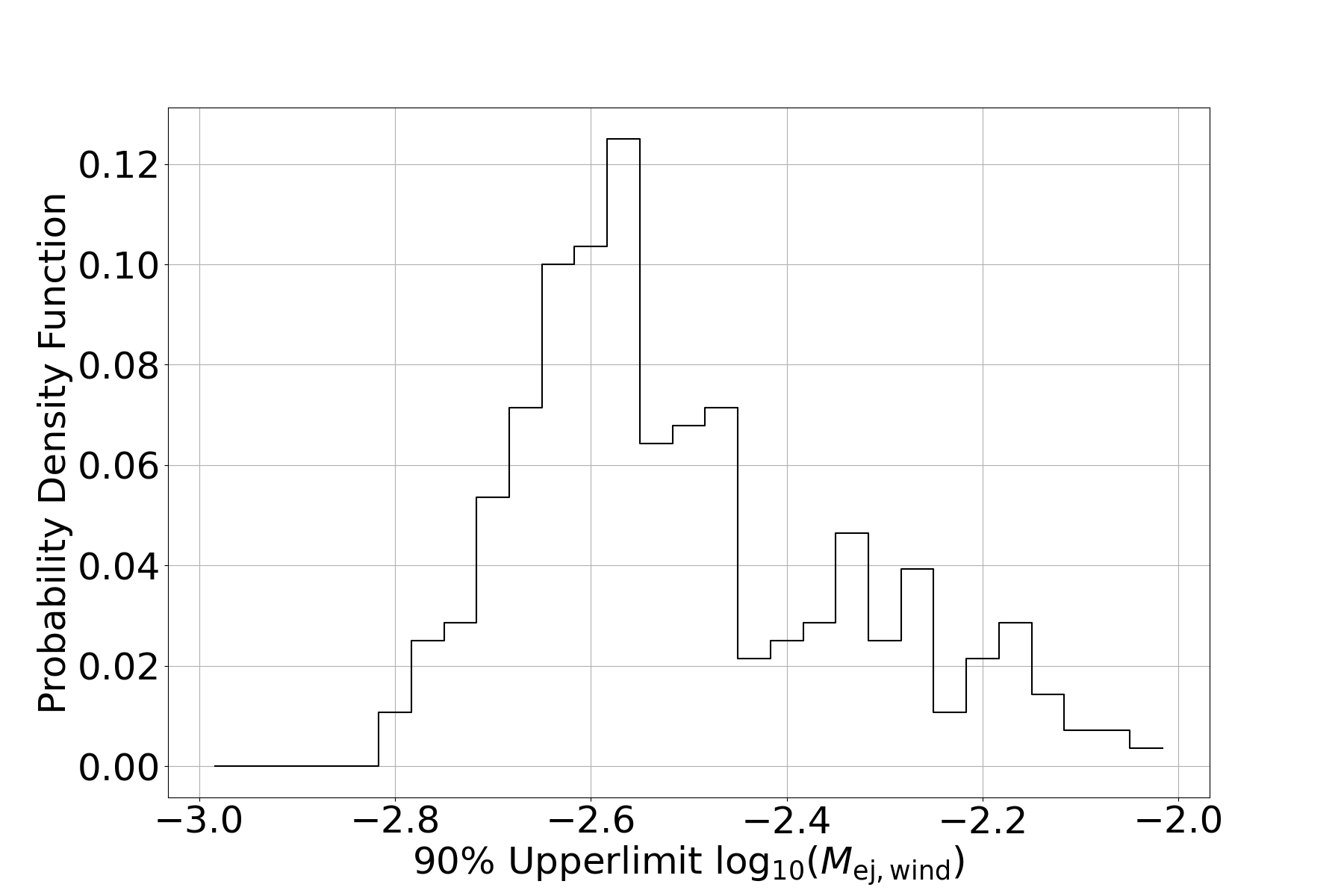}
    \caption{90\% upper limit on the $M_{\rm ej}^{\mathrm{wind}}$ measurements for the \emph{prompt collapse} simulation set, where the $M_{\rm ej}^{\mathrm{wind}}$ contribution is set to 0.
    }
    \label{fig:prompt}
\end{figure}

We also desire to differentiate between the prompt collapse and formation of hypermassive and/or supramassive neutron star.
In the prompt-collapse scenario, where we assume that no disk wind is produced, we show in Figure~\ref{fig:prompt} the 90\% upper limit on the disk wind contribution for our sample. We derive limits ranging from ${\rm log}_{10} M_{\rm ej}^{\mathrm{wind}} = [-2.8 -2.0]$, resulting in strong constraints on the presence of such a disk.

\subsubsection{ZTF proposal for  GW follow-up and triggering criteria}
\label{sec:ZTF-proposal}

The properties of a GW event released in low-latency via the General Coordinates Network (GCN) notice can prove to be useful in determining whether or not to trigger telescope time on a given GW event. Information such as the  alse alarm rate (FAR), probability that the source is of astrophysical origin (p-astro), probability of the GW merger containing a neutron star (HasNS), and probability of the GW merger leaving behind a remnant (HasRemnant), and the Bayes factor of coherence between multiple detectors (log(BCI)) can provide indirect clues about whether a GW event is astrophysical, significant, and could harbor an EM counterpart. Thus, in Table~\ref{tab:ZTF-triggering}, we define triggering criteria for ZTF based on these low-latency properties. An event satisfying all of the ``Go-deep" requirements merits triggering ToO observations, provided the localization and distance are accessible for ZTF. A ``Go-wide" event prompts reweighting public ZTF survey fields to observe the localization in the nominal 30\,s exposures.  Events for which any properties fall within the ``Deliberate" or the ``No Go" categories merit human interaction and discussion to decide whether to trigger. Since the threshold for LIGO-Virgo-KAGRA public alerts release has been lowered to 2 day$^{-1}$ ( user guide\footnote{\url{https://emfollow.docs.ligo.org/userguide/analysis/index.html}}, \citealt{LRR_2020}) in O4, having such triggering criteria will be paramount for wisely allocating existing telescope resources.

\begin{figure*}[ht!]
    \includegraphics[width=7in]{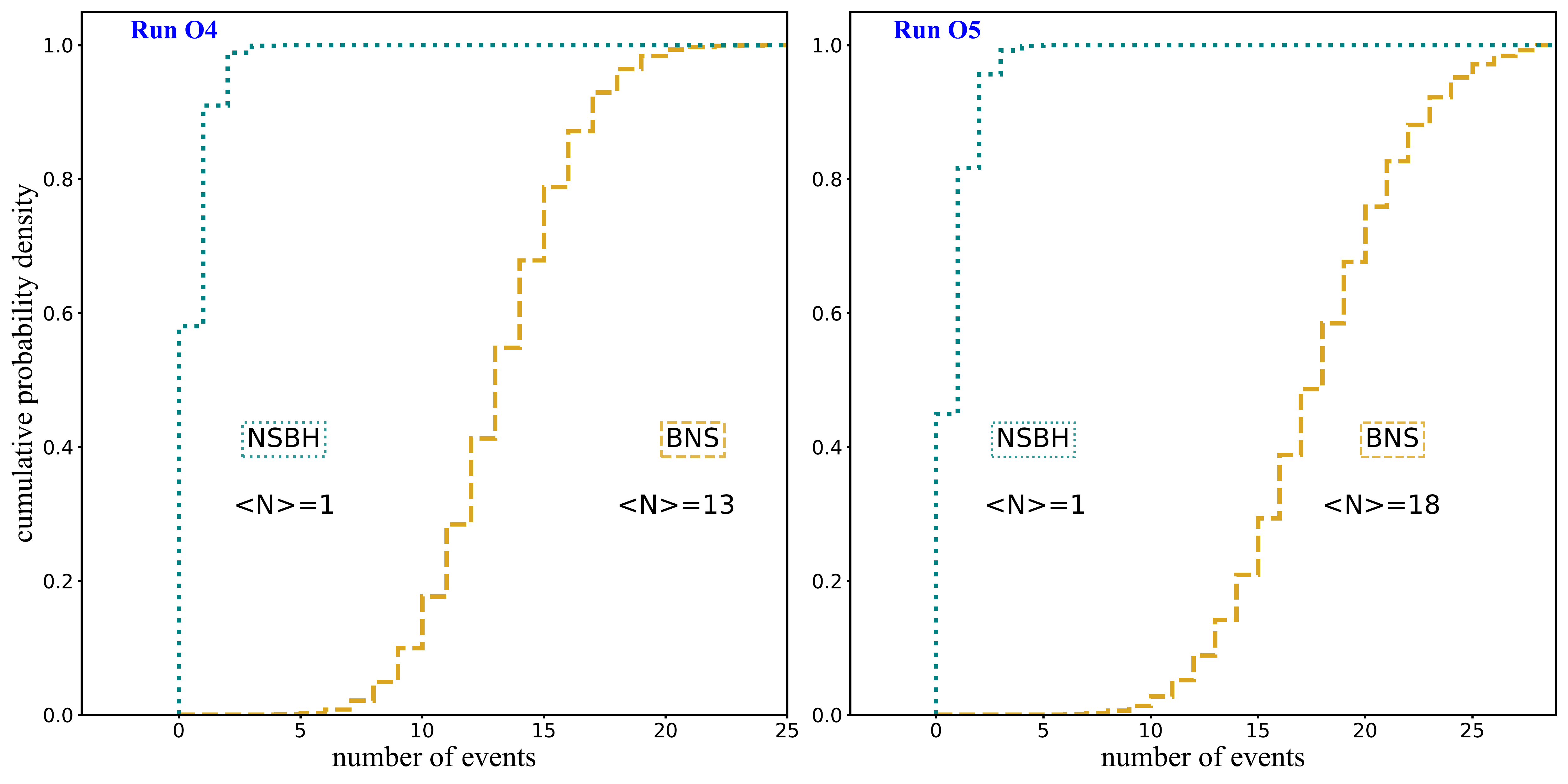}
    \caption{Cumulative histograms of 100,000 realizations of the number of BNS and NSBH mergers predicted to be detected during LIGO-Virgo-KAGRA O4 and O5 within 400\,Mpc based on the observing scenarios predictions in this paper. The mean number of expected detections is quoted for each merger type.
    }
    \label{fig:ztf-detection-proposal}
\end{figure*}

In Figure~\ref{fig:ztf-detection-proposal}, we show the expected annual number of triggers within 400\,Mpc during the O4 and O5 run, based on the predictions in Section~\ref{subsec:observing-scenarios-results}. Assuming a AT2017gfo-like KNe with $M_{\mathrm{abs}}\sim-16$\,mag at peak, and taking to account ZTF's limiting magnitude in 300\,s exposures (m$_{\mathrm{AB}}\approx22$\,mag), we estimate that ZTF could detect KNe falling within 400\,Mpc. Due to the overall low numbers of NSBH mergers, we only expect 0--2 NSBH mergers within 400\,Mpc during the O4 and O5 runs, which is consistent with the KNe detection prospects discussed in Sec.~\ref{sec:kilonovae}. However, our calculations yield $<N>\, =13$ ($<N>\,=18$) BNS mergers within 400\,Mpc during O4 (O5), providing ample opportunities to conduct sensitive searches for counterparts to BNS mergers.

\begin{figure}[ht!]
    \includegraphics[width=3.5in]{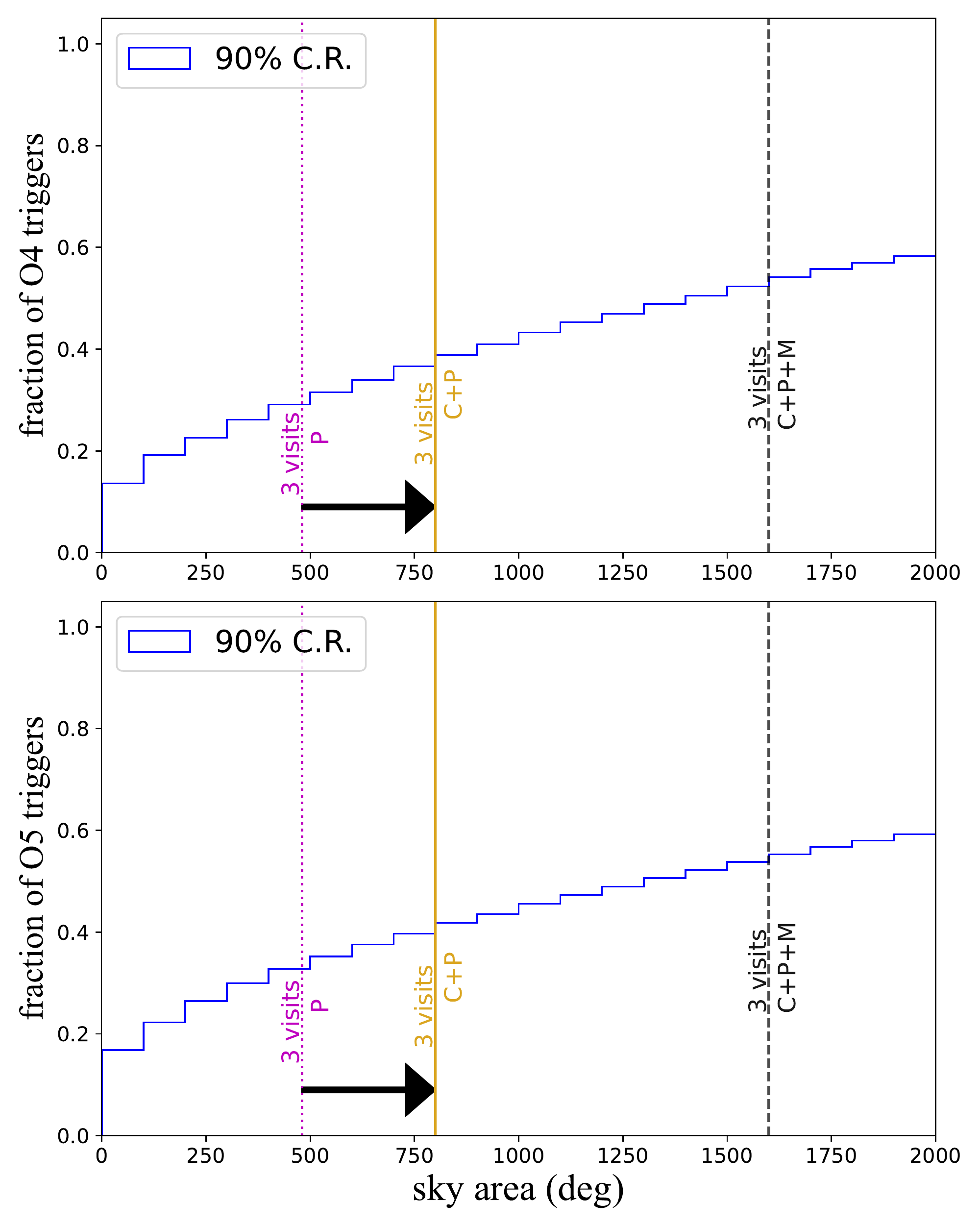}
    \caption{Cumulative histogram of simulated GW skymaps for O4 that satisfy our triggering criteria as a function of the 90\% credible region. With both Caltech and Partnership time spanning $\sim$50\% of the night (corresponding to a maximum area of 800 deg$^{2}$), we could fully probe the localization for nearly 40\% of all events.
    }
    \label{fig:trigger-area}
\end{figure}

Finally, we assess the distribution of GW events as a function of sky area. In Figure~\ref{fig:trigger-area}, we show the fraction of O4/O5 triggers whose 90\% confidence region. of the GW localization falls within a given sky area threshold. Assuming a typical 8\,hr night and ZTF's footprint of $\sim$50 deg${^2}$, and a three-filter tiling strategy (i.e., $g-r-g$) we find that, with ZTF Partnership time alone (comprising 30\% of the night), we can fully tile the localization for $\sim$30\% of GW alerts. With the addition of the private Caltech allocation (comprising $\sim$50\% of the night), we can probe the localization for nearly 40\%, and by using the public survey allocation as well (100\% of the night), we can fully tile the localization for 50\% of all events. These figures of merit can be easily computed from the data set presented in this paper to estimate the number of ToO triggers and time request needed for a successful GW follow-up campaign with wide-field telescopes. In particular, these calculations will prove especially useful for informing Rubin's triggering strategy once it comes online.


\begin{table*}[t]
\renewcommand\arraystretch{1.3}
\setlength{\tabcolsep}{0.43cm}
    \centering
    \caption{Triggering Based on GW Candidate Event Properties}
    \begin{tabular}{c|c|c|c|c}
    \hline\hline
       \textbf{Parameter} &\ \textbf{Go-deep} &\ \textbf{Go-wide} &\ \textbf{Deliberate} &\ \textbf{No Go}\\ 
    \hline\hline
    \multirow{2}*{\textbf{Strategy}}                                   & 300 s                & 30 s              &  \multicolumn{2}{c}{\multirow{4}*{Action Item: human interaction}}     \\
                                                                       & Push distance       & Push localization                                                                             \\
    \cline{1-3}
    \multirow{2}*{\textbf{Frequency of triggers}}       & 1 per month                 & 2 per month                                                                                          \\ 
                                                        & 3 nights                     & 5 nights                                                                                             \\
    \hline 
    \multirow{2}*{\textbf{FAR min(FAR) - ‘Best’}}       & $<$ 1 per century           & $<$ 1 per decade            & 1 per year - century                        & $>$ 1 per year \\
                                                        & Any pipeline                & Any pipeline                &                                             & All pipelines   \\
    \hline
    \textbf{max(p-astro)}                               & $>0.9$                      & $>0.9$                      & 0.1--0.9                                    & $<0.1$          \\
    \hline
    \textbf{HasNS}                                      & $>0.9$                      & $>0.9$                      & 0.1--0.9                                    & $<0.1$           \\
    \hline
    \textbf{log(BCI)}                                   & $>4$                        & $>4$                        & -1 to 4                                     & $<-1$             \\
    \hline
    \textbf{HasRemnant?}                                & $>0$                        & $>0$                        & ...                                         & $=0$               \\
    \hline
    \textbf{pBNS/pNSBH}                                 & $>0$                        & $>0$                        & ...                                         & $=0$                \\
    \hline
    \hline                          
    \end{tabular}
    \label{tab:ZTF-triggering}
\end{table*}

\subsection{$H_0$ constraints}
\label{sec:H0-discussion}

Given the high computational costs for GW analysis runs, we reduce the sample size of events for O4 and O5, and study only the loudest 30 sources in terms of network S/N. This selection of the 30 events was made on the CBCs of \texttt{PBD/GWTC-3} (\texttt{GWTC-3}) distribution that passed the S/N threshold. Therefore, we note that this number is \emph{not} a function of the expected EM detection rate during the next campaigns O4 and O5.
For the GW simulations, we employ the \textsc{IMRPhenomD\_NRTidalv2} model of \cite{Dietrich:2019kaq} and the EOS with maximum likelihood inferred in \cite{Huth:2021bsp}. We inject signals in Gaussian noise, using the detectors' PSD predicted for O4 and O5, as explained above. We perform parameter estimation analyses, where the GW likelihood landscape is explored employing the \texttt{BILBY} library \citep{Ashton:2018jfp,Romero-Shaw:2020owr} and the included nested sampling \citep{Skilling:2006gxv,Veitch:2009hd} algorithm \texttt{DYNESTY} \citep{2020MNRAS.493.3132S,sergey_koposov_2022_6456387}. To reduce the computational cost of the analysis for injected GW signals, we use the relative binning method \citep{Zackay:2018qdy,Leslie:2021ssu}, following the implementation in \cite{janquart:2022}. We analyze signals starting from 20~Hz, and we employ the usual uniform in comoving volume prior to the luminosity distance used in GW parameter estimation. The same prior setting is used for KNe inferences detailed in the following. With regard to potentially associated KNe signals, we use the \texttt{NMMA} framework to infer posterior distributions on KNe properties as well as the luminosity distance using the KNe model of \cite{Bulla:2019muo, Bulla:2022mwo}. Moreover, we perform two sets of EM parameter estimates, for ZTF and Rubin Observatory observations. In order to connect the binary masses in the remaining sample of 30 BNS systems to ejecta material masses powering the KNe, we use the phenomenological relations established in \cite{Dietrich2020}. In this process, there were nine (seven) high-mass BNS systems in the O4 (O5) sample, which would directly form a BH leaving no ejected material powering an EM counterpart.

For the remaining 21 (23) O4 (O5) BNS mergers with an EM counterpart, we use the inferred luminosity distance posterior distributions of both GW simulations and EM simulations to determine the posterior distribution for $H_0$. Moreover, we include inferred luminosity distance posteriors of the GW signal GW170817  from \cite{LIGOScientific:2018hze},  \cite{LIGOScientific:2018mvr} and the KN signal AT2017gfo, which we inferred with the same KN model yielding a total number of 22 (24) O4 (O5) BNS coalescences. In this context, we assume the standard cosmology model. Since all of our BNS systems lie within a range of $300$~Mpc or within a small redshift regime, we assume the appropriate linear Hubble relation for nearby events
\begin{equation}\label{Eq:H0}
    c z \approx H_0 d_L,
\end{equation}
in which $d_L$ and $z$ are the luminosity distance and the redshift to the source, respectively, and $c$ is the speed of light, and $H_0$ is the Hubble constant. Since a volumetric prior on the luminosity distance inherits a $1/H_0^4$ prior factor, we correct for this in our study similar to \cite{LIGOScientific:2017adf}. 
Concerning the underlying distance and/or redshift distribution of our injected O4 (O5) samples, we use the injected distance and the injected Hubble constant to calculate the corresponding injected redshift, which is the mean of the Gaussian distribution with a 1\% relative uncertainty set as standard deviation. In this study, we do not break the distance-inclination angle degeneracy by including additional information on the binary's orbital inclination from other potential observations, such as GRBs. Nevertheless, we are able to improve the Hubble constant measurements through the combined distance measurements from GW and KNe observations; see Figure~(2a) for a similar approach taken in \cite{Dietrich:2020efo}.

We point out that, for a more detailed study with a larger number of considered detections, further selection effects would need to be considered for our population analysis. In particular, we would need to correct for our choice that we consider only the loudest 30 GW signals for which we simulated potential KNe observations. Apart from a selection effect that would result from the redshift measurement of a potential host galaxy, we would also need to consider the dependency of $H_0$ on the BNS component masses as high-mass BNS systems will quickly form as BHs leaving no traceable EM counterpart and, consequently, were not considered for our $H_0$ projections. We will leave a detailed analysis of selection biases for future work at the present moment. However, although additional selection bias corrections might be needed, our analysis does not show biases in our recovery, which overall shows the robustness of our study, but we expect this situation to change when the population sample sizes increase or observations at higher redshifts are included.

\begin{figure*}
    \includegraphics[width=3.5in]{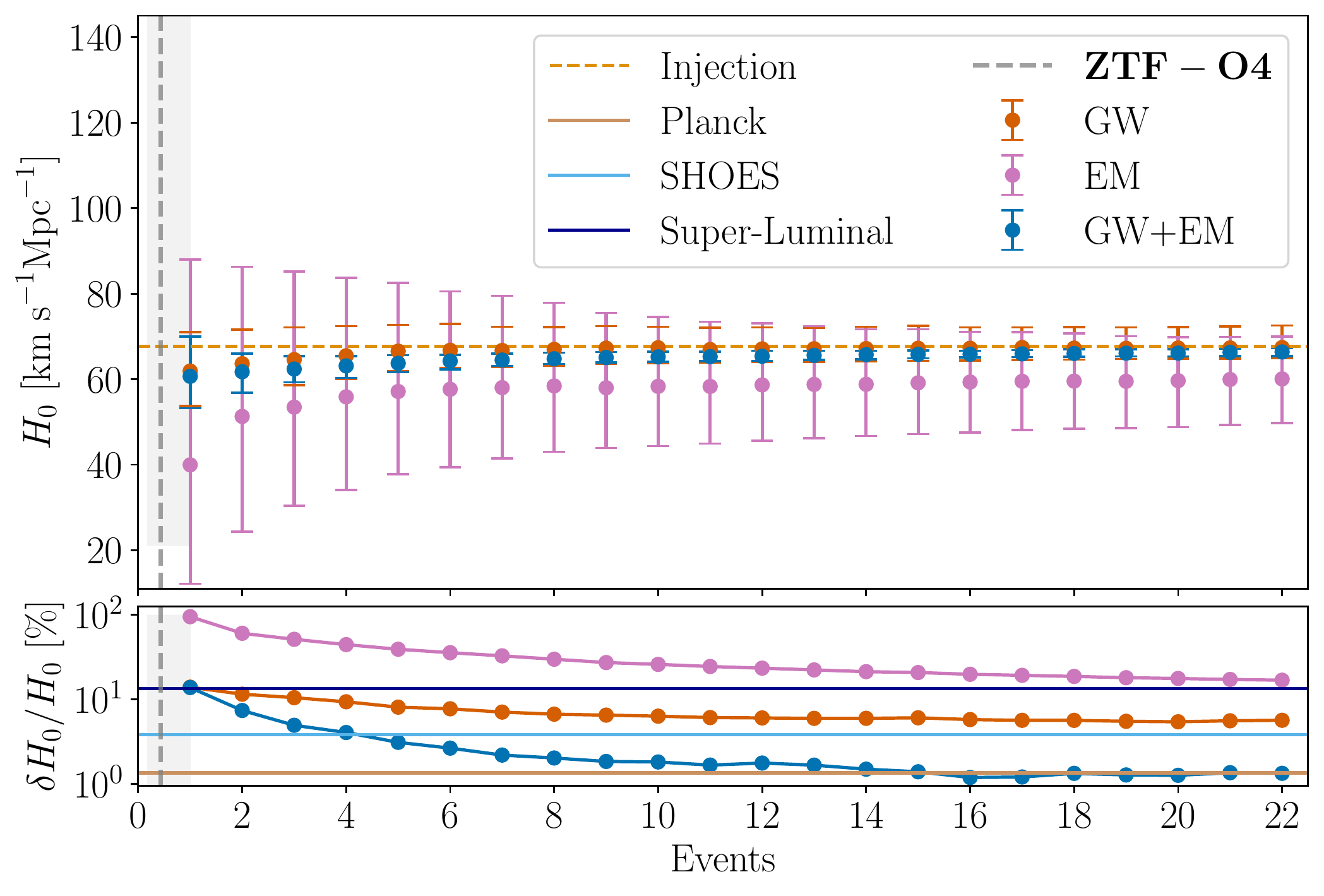}
    \includegraphics[width=3.5in]{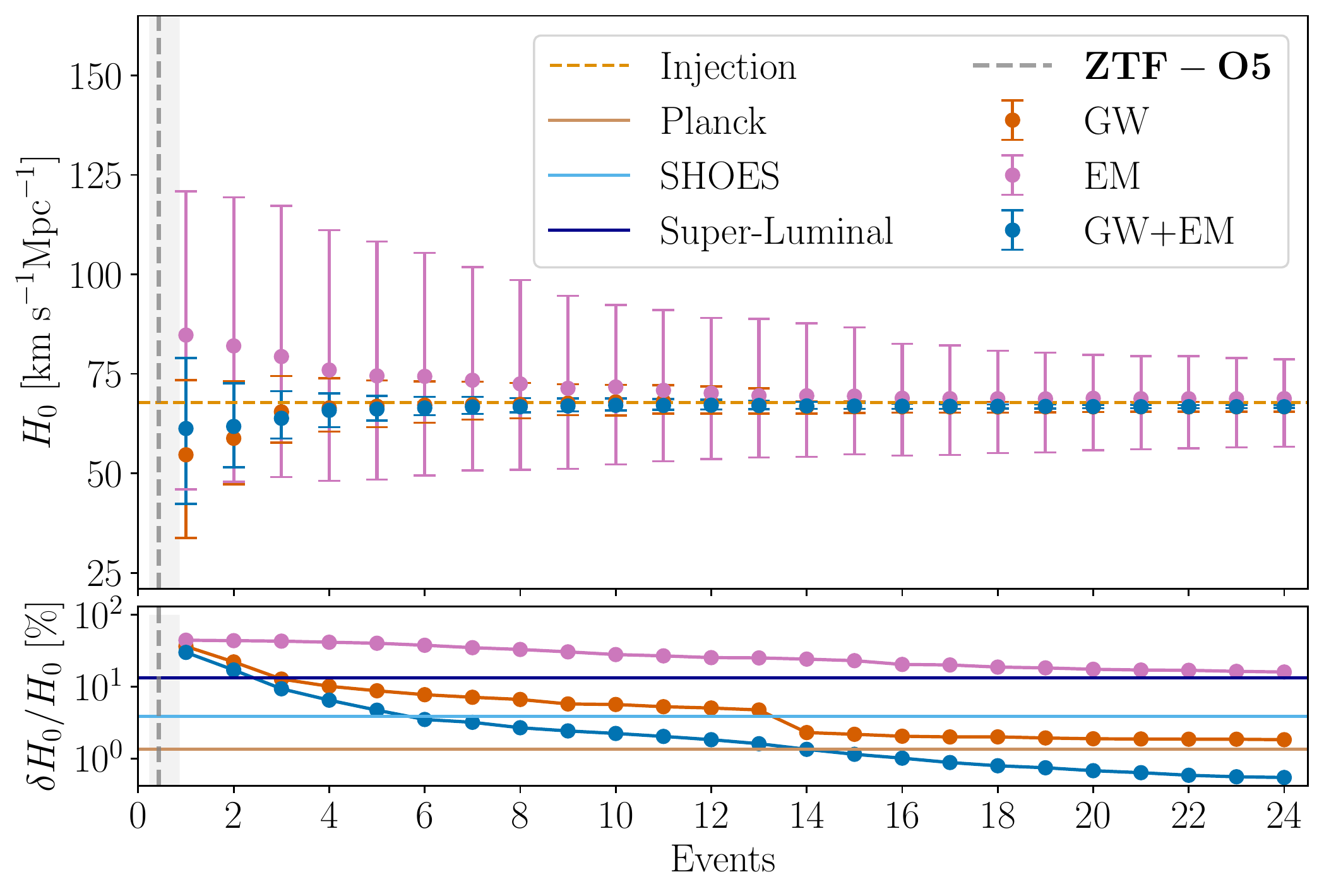}
    \includegraphics[width=3.5in]{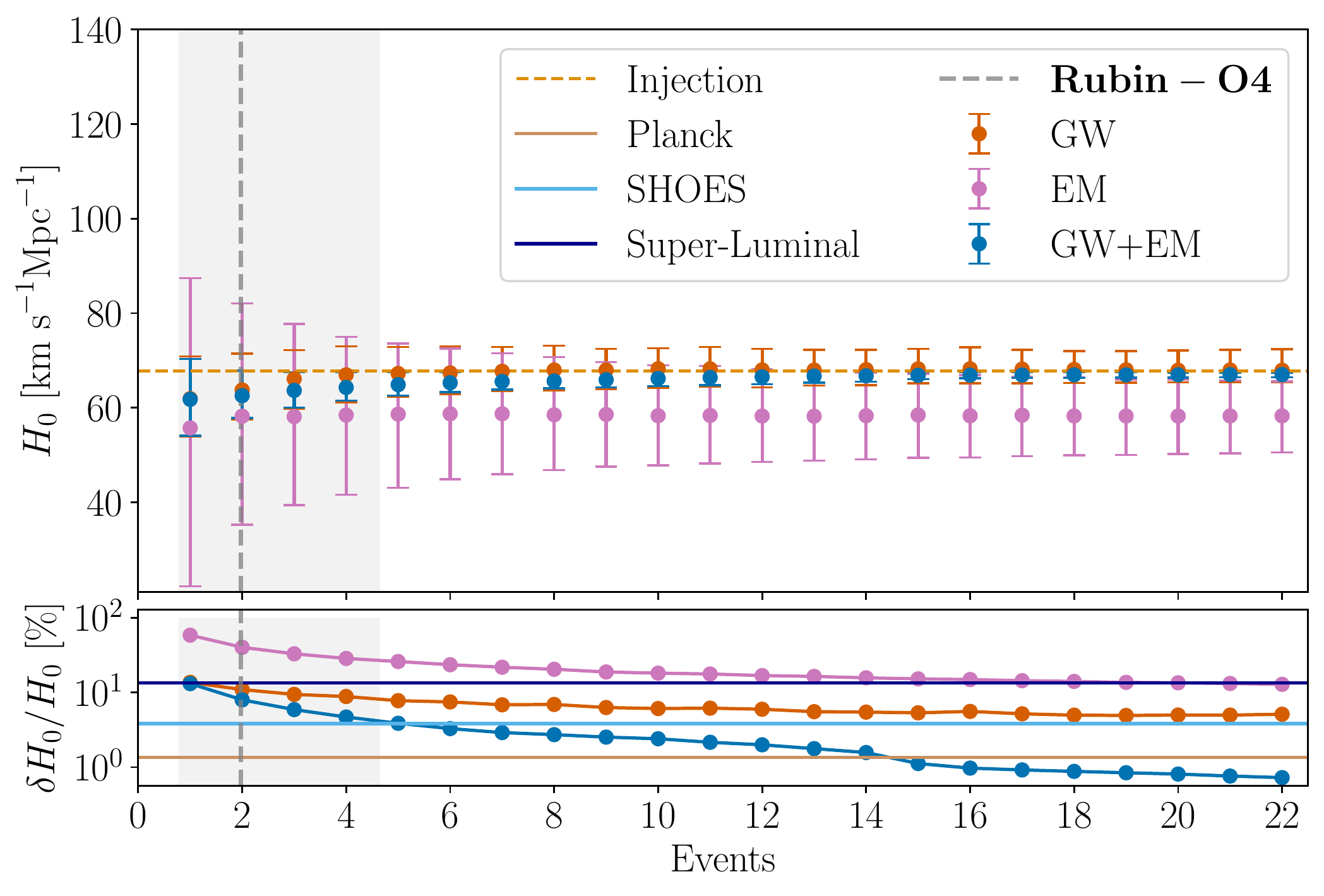}
    \includegraphics[width=3.5in]{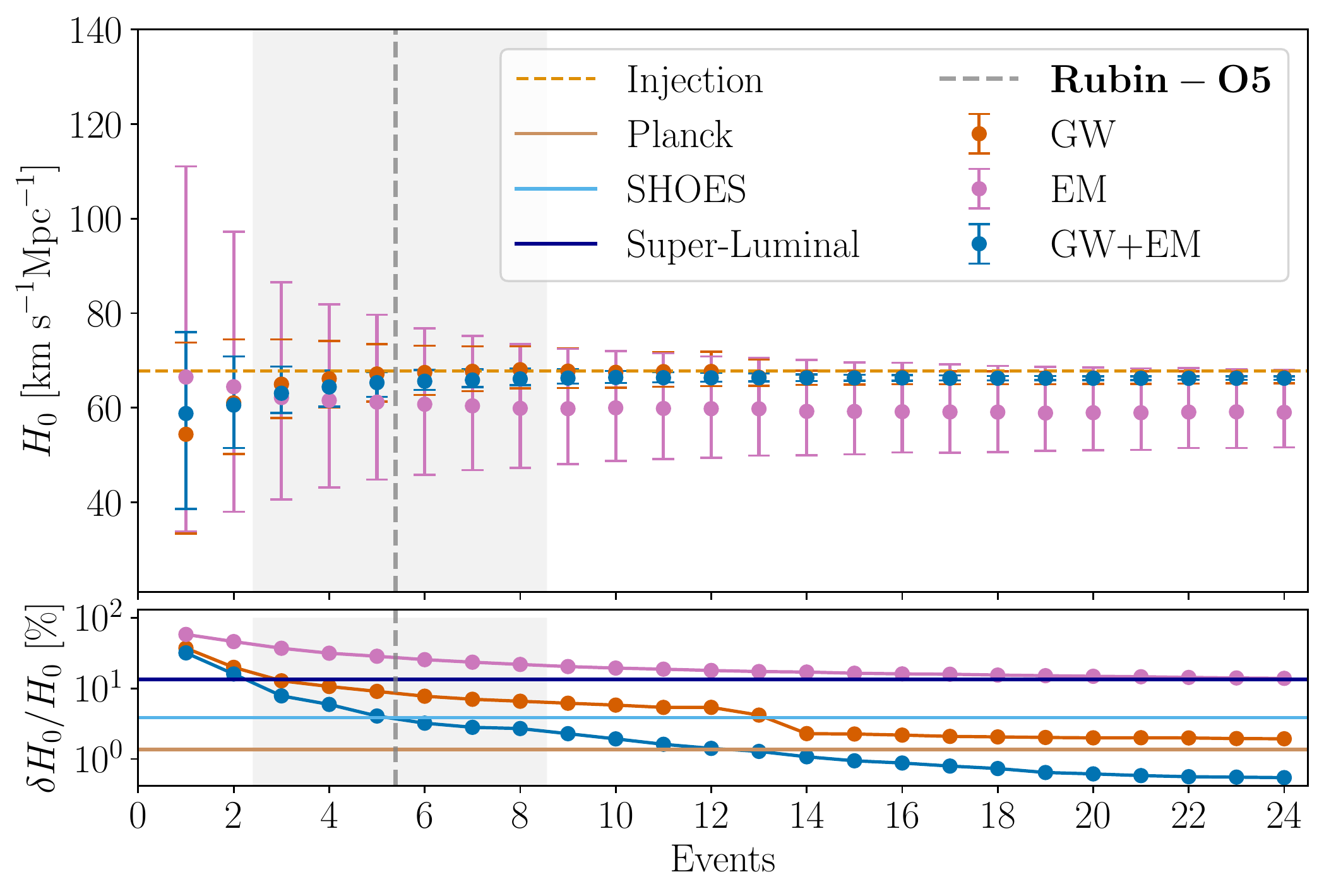}
    \caption{
    ${H_0}$-estimates for the ZTF (top row) and Rubin Observatory (bottom row) observation scenarios for O4 (left column) and O5 BNS samples (right column). We show ${H_0}$-estimates from our GW simulations (in orange), EM simulations (in violet), and for the combined GW+EM results (in blue) in the top panels, whereas relative errors are shown in the bottom panels. We indicate the expected O4 and O5 detection rates in alignment with Table~\ref{tab:em-annual-detection-rate} as gray dashed line and show  the 90\% credible interval as gray shaded regions. 
    In the bottom panel, we contrast our results to the Planck measurement of the  cosmic microwave background (\citealt{Ade:2015xua}; Planck, violet), to the Hubble measurement via type-Ia supernovae (\citealt{Riess:2016jrr};SHOES, light blue), and to the ${H_0}$-measurement of superluminal motion of the jet in GW170817 (\citealt{Hotokezaka:2018dfi}; superluminal, gray). All uncertainties are reported at 90\% credible interval.
    }
    \label{fig:H0}
\end{figure*}

In Figure~\ref{fig:H0} (top panels), we show our $H_0$-results for the ZTF scenario obtained from GW and EM parameter estimations on the luminosity distance and, in addition, the results when combining GW+EM information and contrast these to state-of-the-art measurements. The uncertainties of our results are reported at 90\% credible interval. We highlight our estimated annual BNS detection rate from Table~\ref{tab:em-annual-detection-rate} as gray dashed line and mark the corresponding 90\% credible interval as gray regions. With only one joint GW+EM observation, as estimated for ZTF-O4, we are not able to provide strong constraints on the Hubble constant, i.e., while we recover the injection value of $H_0 =$ 67.74~$\rm{km \,s^{-1} Mpc^{-1}}$; this is mainly caused by the large uncertainty of our measurement $H_0 = 60.68^{+9.24}_{-7.47}~\rm{km \,s^{-1} Mpc^{-1}}$. Most notably, this shows how unlikely it is that within O4 we will be able to break the Hubble tension. Similarly, one GW+EM observation in O5 provides an estimate of $H_0 = 61.26^{+17.73}_{-18.97}~\rm{km \,s^{-1} Mpc^{-1}}$. To show what might be possible with several more GW events that have a corresponding KNe and to understand whether there is a systematic bias being introduced with our methodology, we combine 22 BNS events in O4 and estimate $H_0 = 66.37^{+0.58}_{-0.95}~\rm{km \,s^{-1} Mpc^{-1}}$.
We do a similar study for O5, combining 23 events, and estimate $H_0= 66.74^{+0.39}_{-0.33}~\rm{km \,s^{-1} Mpc^{-1}}$.
Both estimates recover the injected value of $H_0$.

In Figure~\ref{fig:H0} (bottom panels), we show our $H_0$-results for the Rubin Observatory scenario. We find that roughly two joint observations as estimated for O4 with $H_0 = 62.56^{+5.27}_{-4.70}~\rm{km \,s^{-1} Mpc^{-1}}$ and approximately 5 combined observations in O5 with $H_0 = 65.30^{+2.31}_{-2.99}~\rm{km \,s^{-1} Mpc^{-1}}$ can recover the injection. For a total number of 22 combined events in O4, we estimate $H_0=67.01^{+0.43}_{-0.53}~\rm{km \,s^{-1} Mpc^{-1}}$ which is close to the injection value. For O5, we recover $H_0= 66.23^{+0.39}_{-0.33} ~\rm{km \,s^{-1} Mpc^{-1}}$. 
Overall, our study shows that subpercent level measurements will be possible through a combination of GW+EM (KNe) information with a sufficient number of KNe detections. Furthermore, we emphasize that, while we include only the distance measurement from the KNe, further information, e.g., through the GRB afterglow might further help to reduce uncertainties by breaking degeneracies between the distance and the inclination.

\section{Conclusion}
\label{sec:conclusion}    

In this paper, we have performed an end-to-end GW survey simulation for O4 and O5 follow-up of BNS and NSBH mergers.
We simulate event follow-up for the ZTF and Rubin Observatory, showing the impact on their detection capability. 
Based on the GW and EM posteriors from these analyses, we simulate the potential $H_0$  using a distance threshold of $300$~Mpc.
With the simulation study performed above, we can summarize our conclusions and implications for future GW follow-ups as follows:

\begin{enumerate}[label=(\roman*)]
\item During O4 and O5, there will be many more events than are reasonable to follow up. Unsurprisingly, the objects that are best to follow-up are the best localized and near-by \cite{ChHo2016}. Proximity is an especially important consideration for the NSBH mergers, given their large expected distances and faint intrinsic luminosities.
\item During the next O4 and O5 runs, in contrast to Rubin Observatory,  ZTF, due to shallower limits in its bands, will have difficulty detecting the EM counterparts, especially at larger distances (see Figure~\ref{fig:detection-O4}).
\item  The GW contribution dominates under the assumptions in this document, but again, the high S/N, nearby events dominate the sensitivity improvement. A few well-localized nearby events are worth much more than many far-away, poorly localized events. For $H_0$, most events contribute equally, especially as relative uncertainty due to virial velocities decreases as distance increases. 
\end{enumerate}

\section{Acknowledgements}

M.W.C acknowledges support from the National Science Foundation with grant numbers PHY-2308862 and OAC-2117997.
M.W.C and R.W.K were supported by the Preparing for Astrophysics with LSST Program, funded by the Heising Simons Foundation through grant 2021-2975, and administered by Las Cumbres Observatory.
A.M.F is supported by the National Science Foundation Graduate Research Fellowship Program under grant No. DGE-1746045.

The authors are very grateful to Sharan Banagiri for his review on behalf of P\&P, which has been useful  in the improving of this paper.
We  would like to thank the Cosmology GW group and in particular  Rachel Gray  and Thomas Dent for the suggestions and comments that allowed us to improve our paper.

The authors acknowledge the International Gravitational-wave Network  Computing Grid (CIT, LHO, LLO) for providing resources to realize our simulations.
We acknowledge the Minnesota Supercomputing Institute (MSI; \url{http://www.msi.umn.edu})  at the University of Minnesota for providing resources that contributed to the research results reported within this paper under project ``Identification of Variable Objects in the Zwicky Transient Facility.''
This research used resources of the National Energy Research Scientific Computing Center (NERSC), a U.S. Department of Energy Office of Science User Facility operated under contract No. DE-AC02-05CH11231 under project ``Toward a complete catalog of variable sources to support efficient searches for compact binary mergers and their products.''
This work used the Advanced Cyberinfrastructure Coordination Ecosystem: Services \& Support (ACCESS). This work used the Extreme Science and Engineering Discovery Environment (XSEDE) COMET at SDSU through allocation AST200016 and AST200029. 
This material is based upon work supported by NSF's LIGO Laboratory, which is a major facility fully funded by the National Science Foundation



\bibliographystyle{aasjournal}
\bibliography{references} 

\clearpage

\appendix

\section{ Observing Scenarios}
\label{App:appendix_A}

\subsection{Values of hyperparameters}\label{App:Values_of_Hyperparam}

We fix the parameters in both of our assumed models for the astrophysical distribution of compact binary mergers.
These population model parameters (termed ``hyperparameters'' when used in a hierarchical Bayesian analysis) are listed in Table~\ref{tab:hyperpe-vals} for the \texttt{PBD/GWTC-3} model and were chosen because they correspond to the maximum value of the hyperposterior obtained by fitting the PDB model to GWTC-3 \citep{GWTC-3_2023}.
We are therefore not considering the full posterior uncertainty in these hyperparameters.
To estimate the effect of neglecting this uncertainty on the number of detected events in each subpopulation, we compare the uncertainty astrophysical merger rate predicted the full PDB analysis presented in \citealt{GWTC-3_2023} to those assumed in this work.
The percent uncertainty ($\frac{90\% \text{CI}}{\text{mean}}$ ) in Table ~\ref{tab:merger-rate-density} for the PDB/GWTC-3 model is $171\%$ each for BNS, NSBH, and BBH.
The percent error reported by \cite{GWTC-3_2023} is 229\%, 178\%, and 68\%  for BNS, NSBH, and BBH events, respectively.
We therefore conclude that we underestimate the uncertainty in merger rate BNS and NSBH by factors of 1.34 and 1.04, respectively, and we overestimate the errors for the BBH merger rate by a factor of 2.5.
This is likely because the shape of the BBH mass spectrum is well-measured compared to that of the NS-containing events, making the extra uncertainty introduced by fitting the hyperparameters relatively small.
The opposite is true for NSBHs and BNSs since only $\sim4$ events are used to constrain the shape of the low-mass end of the distribution.

Naively translating the additional uncertainty in the astrophysical merger rate to an uncertainty in the annual number of detected events in O4 yields an increase in uncertainty of 12.2 events for BNS, an increase in uncertainty of 0.24 events for NSBH, and a decrease in uncertainty of 104 events for BBH, assuming symmetric errors.
These corrections are approximate and are meant to give an estimate of the effect of neglecting the uncertainty in population hyperparameters under the \texttt{PBD/GWTC-3} model.
A similar estimate is not available for the \texttt{LRR} model, as the authors do not know of a population fit performed using that framework on GWTC-3 data.

The difference between the maximum a posteriori underlying population and the full underlying population hyperposterior can be seen in Figure 5 of \cite{Farah_2022}.

\begin{table*}[h!]

    \centering
        \caption{
    Summary of Population Model parameters. The first several entries describe the rate and mass distribution parameters, and the last two entries describe the spin distribution parameters. 
    }   
    \begin{tabular}{ c  p{11cm} p{2mm} p{3cm} }
        \hline \hline
        \textbf{ Parameter} & \textbf{Description} &  & \textbf{Value} 
        \\ \hline
        $\alpha_1$ & Spectral index for the power law of the mass distribution at low mass. &  & -2.16 \\ 
        $\alpha_2$ & Spectral index for the power law of the mass distribution at high mass. &  & -1.46 \\ 
        $\mathrm{A}$ & Lower-mass gap depth. &  & 0.97 \\
        $M^{\mathrm{gap}}_{\rm low}$ & Location of lower end of the mass gap. &  & $2.72\, {M}_\odot$ \\
        $M^{\mathrm{gap}}_{\rm high}$ & Location of upper end of the mass gap. &  & $6.13\, {M}_\odot$ \\
        $\eta_{\rm low}$ & Parameter controlling how the rate tapers at the low end of the mass gap. &  & 50 \\ 
        $\eta_{\rm high}$ & Parameter controlling how the rate tapers at the high end of the mass gap. &  & 50 \\ 
        $\eta_{\text{min}}$ & Parameter controlling tapering of the power law at low mass. &  & 50 \\
        $\eta_{\text{max}}$ & Parameter controlling tapering of the power law at high mass. &  & 4.91 \\
        $\beta$ & Spectral index for the power-law-in-mass-ratio pairing function. &  & 1.89 \\ 
        $M_{\rm min}$ & Minimum mass of the mass distribution. &  & $1.16\, {M}_\odot$ \\
        $M_{\rm max}$ &  Onset location of high-mass tapering. &  & $54.38\, {M}_\odot$ \\
        $a_{\mathrm{max, NS}}$ &  Maximum allowed component spin for objects with mass  $< 2.5\, {M}_\odot$.  &  & $0.4$\\ 
        $a_{\mathrm{max, BH}}$ &  Maximum allowed component spin for objects with mass $\geq 2.5\, {M}_\odot$. &  & $1$\\ 
        \hline\hline
    \end{tabular}
  \label{tab:hyperpe-vals}
\end{table*}

\newpage

\subsubsection{Simulated mass distributions for O5}\label{App:pop-dist-O5}

\begin{figure*}[h!]
    \centering
        \includegraphics[scale=0.5]{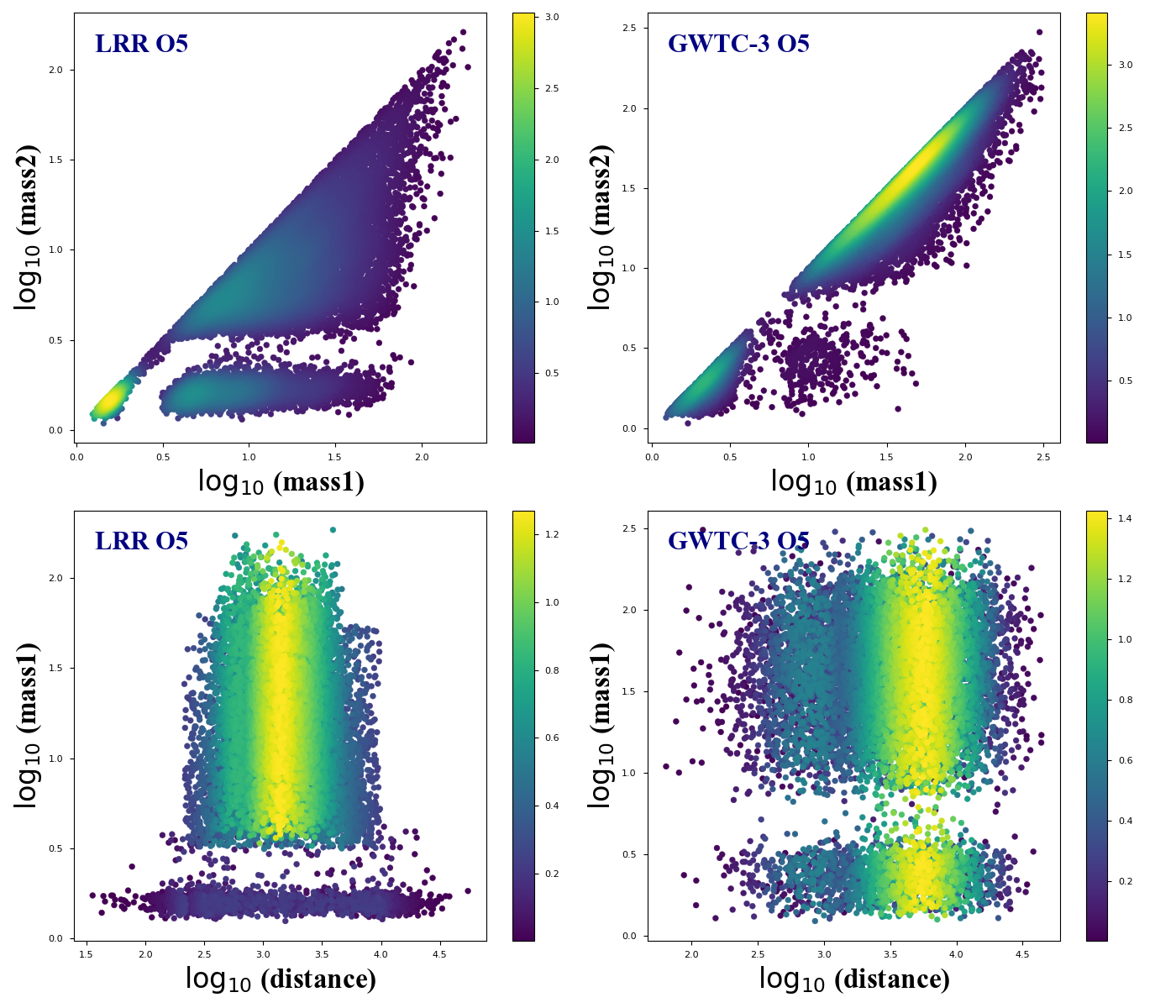}
    \caption{Here we show all the O5 CBCs from all populations. On the left the distribution of \texttt{LRR}, on the right that of \texttt{PBD/GWTC-3} (\texttt{GWTC-3}). The high panels are the mass distributions of the components of each CBC in the context of the detector and the low panels are the primary mass distributions as a function of the collision distance. Mass and distance distributions are shown on a logarithmic scale. The color base shows the number of CBC events per pixel.}
    \label{fig:pop-dist-O5}
\end{figure*}

Figure~\ref{fig:pop-dist-O5} describes the simulated O5 mass distributions for
each model that meets the signal-to-noise (S/N) threshold outlined in Section ~\ref{subsec:simulation}.

\newpage

\subsubsection{Histogram of simulated detections for O5.}\label{App:detection-number-O5}

\begin{figure*}[h!]
    \centering
    \includegraphics[scale=0.5]{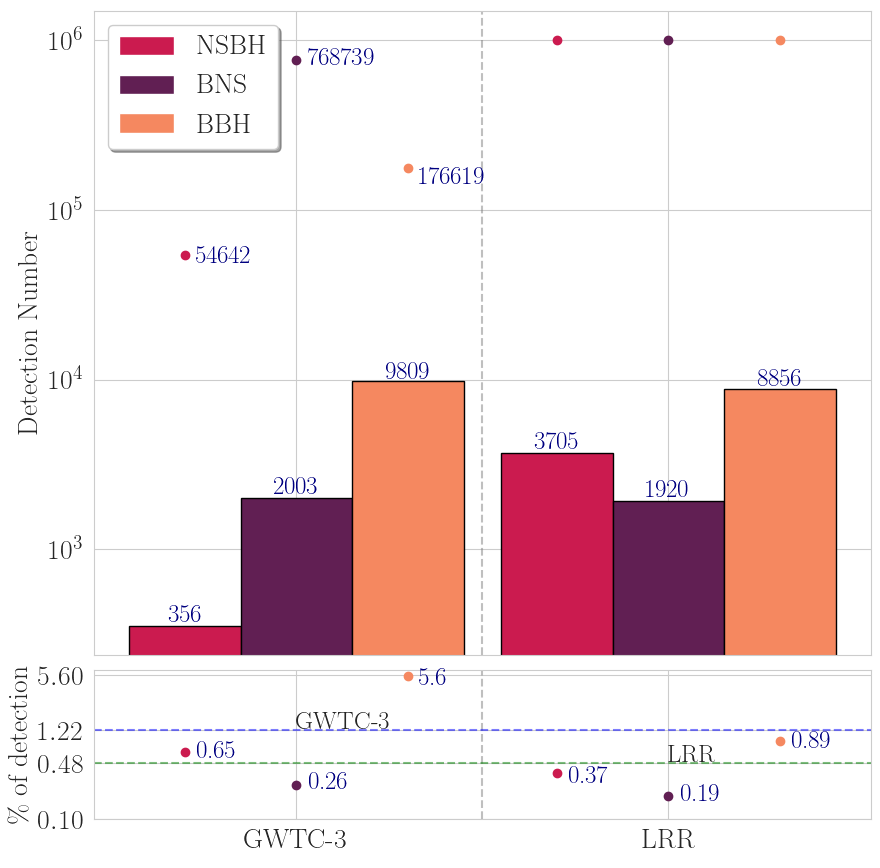}
    \caption{
         This figure shows the simulated detections for O5. The upper panel shows 
         the number of injections (CBCs from each 
         population drawn from both 
         distributions) in colored dots and the 
         bar chart represents the number of 
         events passing the cut-off point. The 
         bottom panel shows the percentage of detections relative to the number 
         injections for the two distributions. The colored dots represent the percentages of each population that passed the S/N,
            while the blue and green lines with respectively $1.22\%$ and $0.48\%$ are successively percentages of detection of all the events (BNS + NSBH + BBH) of the 
         \texttt{PBD/GWTC-3} (\texttt{GWTC-3}) and \texttt{LRR} distributions injected inn our simulation.}
    \label{fig:detection-number-O5}
\end{figure*}

Figure ~\ref{fig:detection-number-O5} we summarize the O5 detection results of the simulation set.

\section{Predictions for detection rates and science with gravitational-wave counterparts}

\subsection{2D histogram of the O5 BNS  and NSBH light curves simulated.}\label{App:lightcurves-histogram-O5}
\begin{figure*}[ht!]
\begin{center}
    
\centering
    \includegraphics[width=3.5in,height=4in]{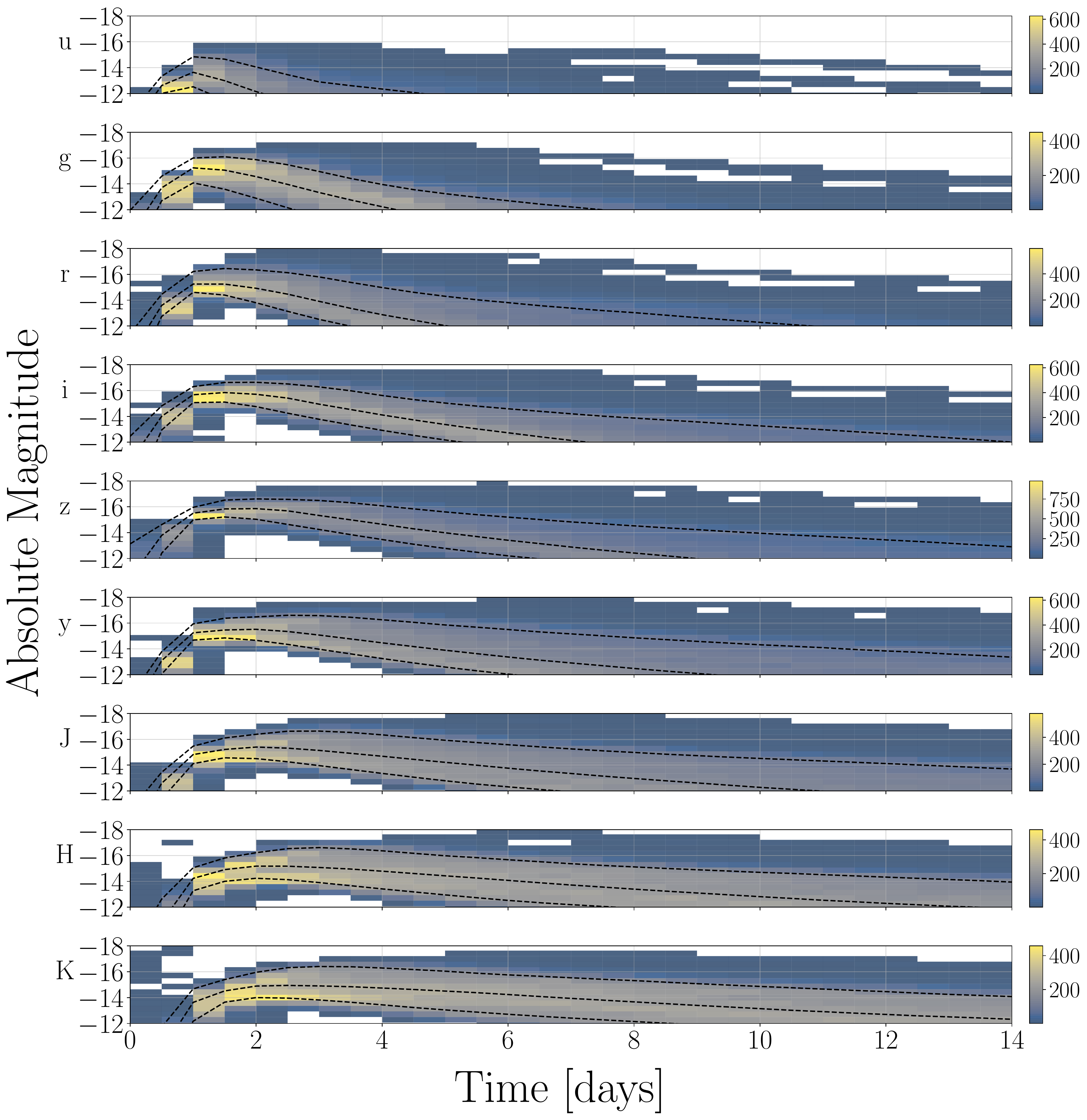}
    \hspace{0.06cm}
    \includegraphics[width=3.5in, height=4in]{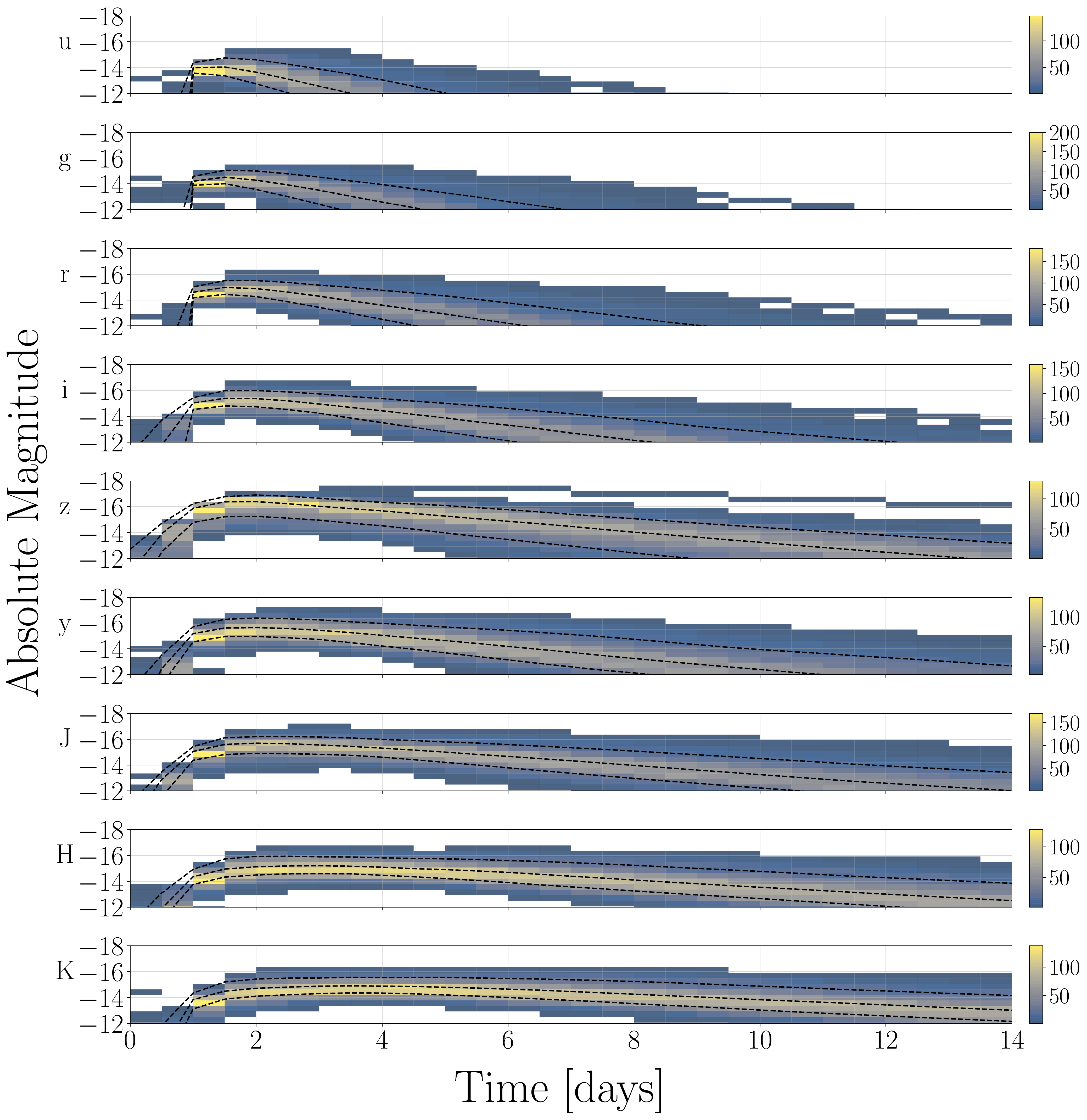}
    \caption{ 2-D histograms of simulated BNS (left) and NSBH (right) light curves for O5. We show light curves in $u$, $g$, $r$, $i$, $z$, $y$, $J$, $H$ bands
             in order to include all the bands used by surveys considered in this work (ZTF, Rubin, and Gemini). 
             In each panel, we plot three dashed lines; on top the 10th percentile, 
             at middle the 50th percentile and on bottom the 90th percentile. The color bar shows the number of detections in the different bands.
            }
    \label{fig:lightcurves-O5}

    \end{center}
\end{figure*}

Here, Figure~\ref{fig:lightcurves-O5} shows a histogram of the light curves in the optical to near-infrared bands for the O5 simulation set, simulated with NMMA and publicly available on \url{https://github.com/nuclear-multimessenger-astronomy/nmma}.

\subsection{Histograms of peak magnitudes}\label{App:peak-mag}
\begin{figure*}[ht!]
\centering
    \includegraphics[width=7.4in]{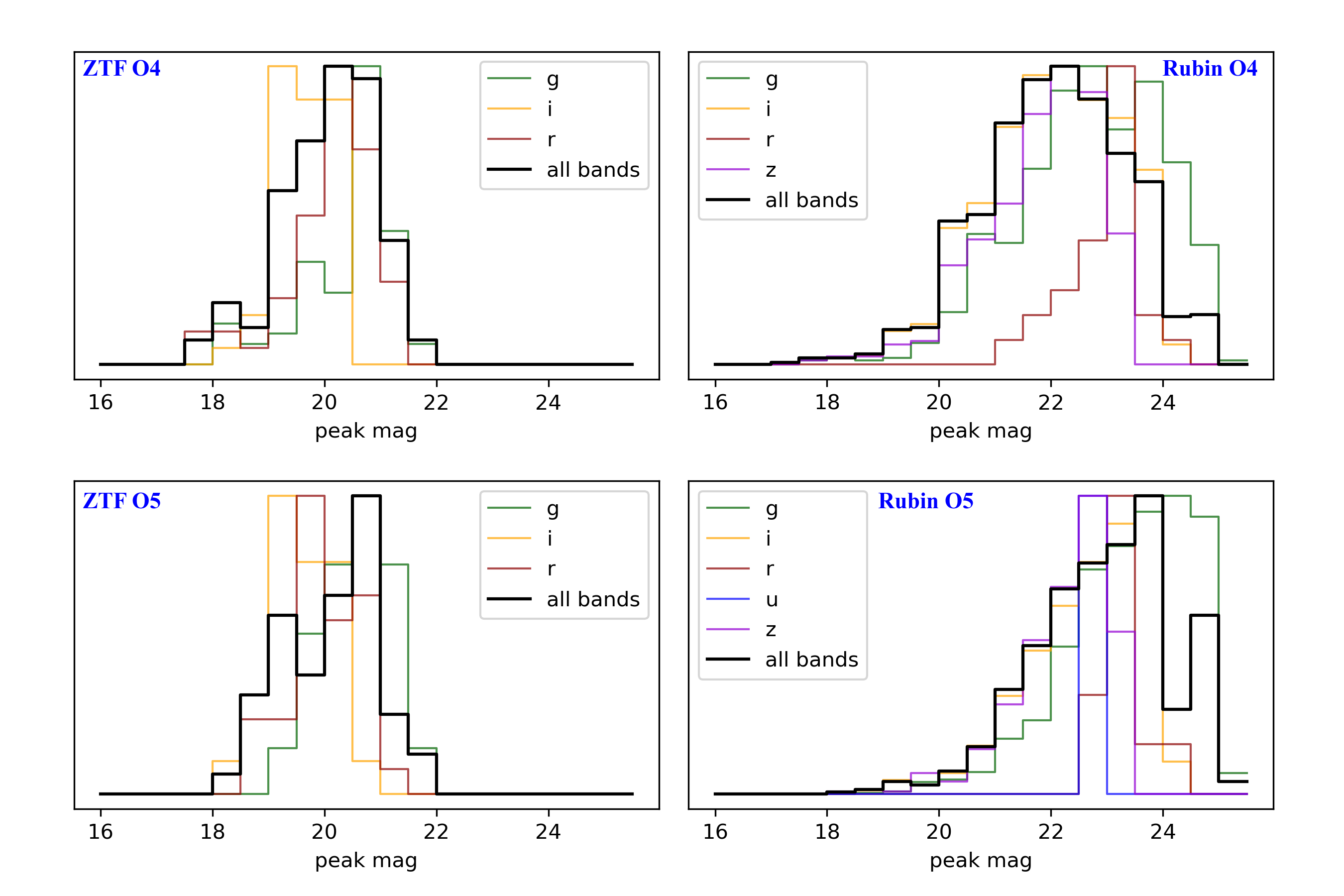}
    \caption{ Those  plots are the 1D histograms of the peak magnitudes in the ZTF  and the Rubin observatory bands. On the left we show  ZTF  BNS peak mag in the run O4 (on top) and O5( at bottom), then on the right the same plot for the case of  Rubin Observatory. The black line shows the peak mag in all the bands of each telescope detected during the run O4 and O5 simulation.)}
    \label{fig:peak_mag_in_each_band}
\end{figure*}

Figure \ref{fig:peak_mag_in_each_band} we present a one-dimensional histogram of predictions for the peaks related to each band of ZTF and the Rubin Observatory during the forthcoming observing runs O4 and O5.
 
\subsection{Distribution of missed and found BNS and NSBH injections by ZTF and Rubin Observatory}\label{App:detection}

\begin{figure*}[ht!]
    \includegraphics[width=3.5in]{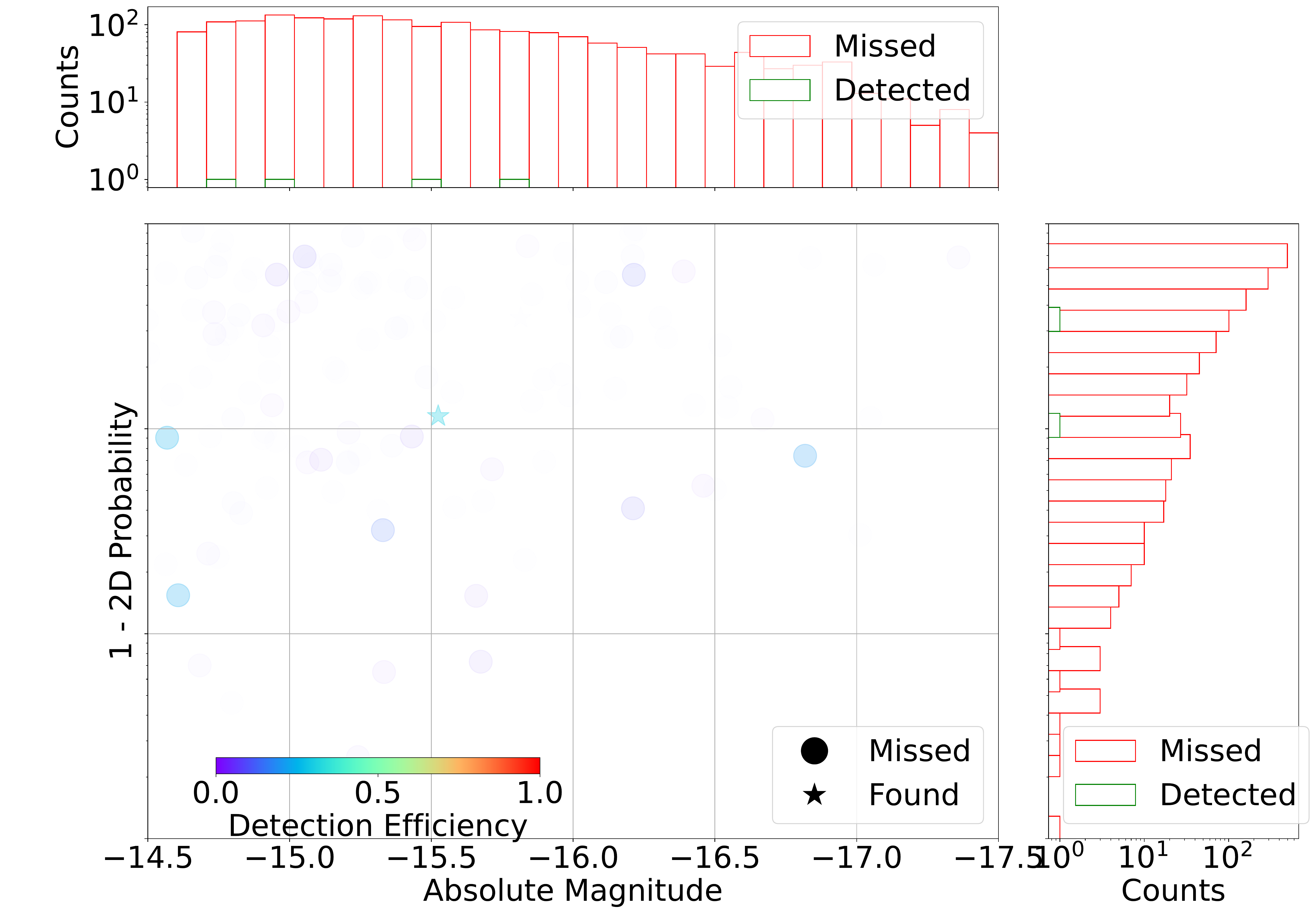}
    \includegraphics[width=3.5in]{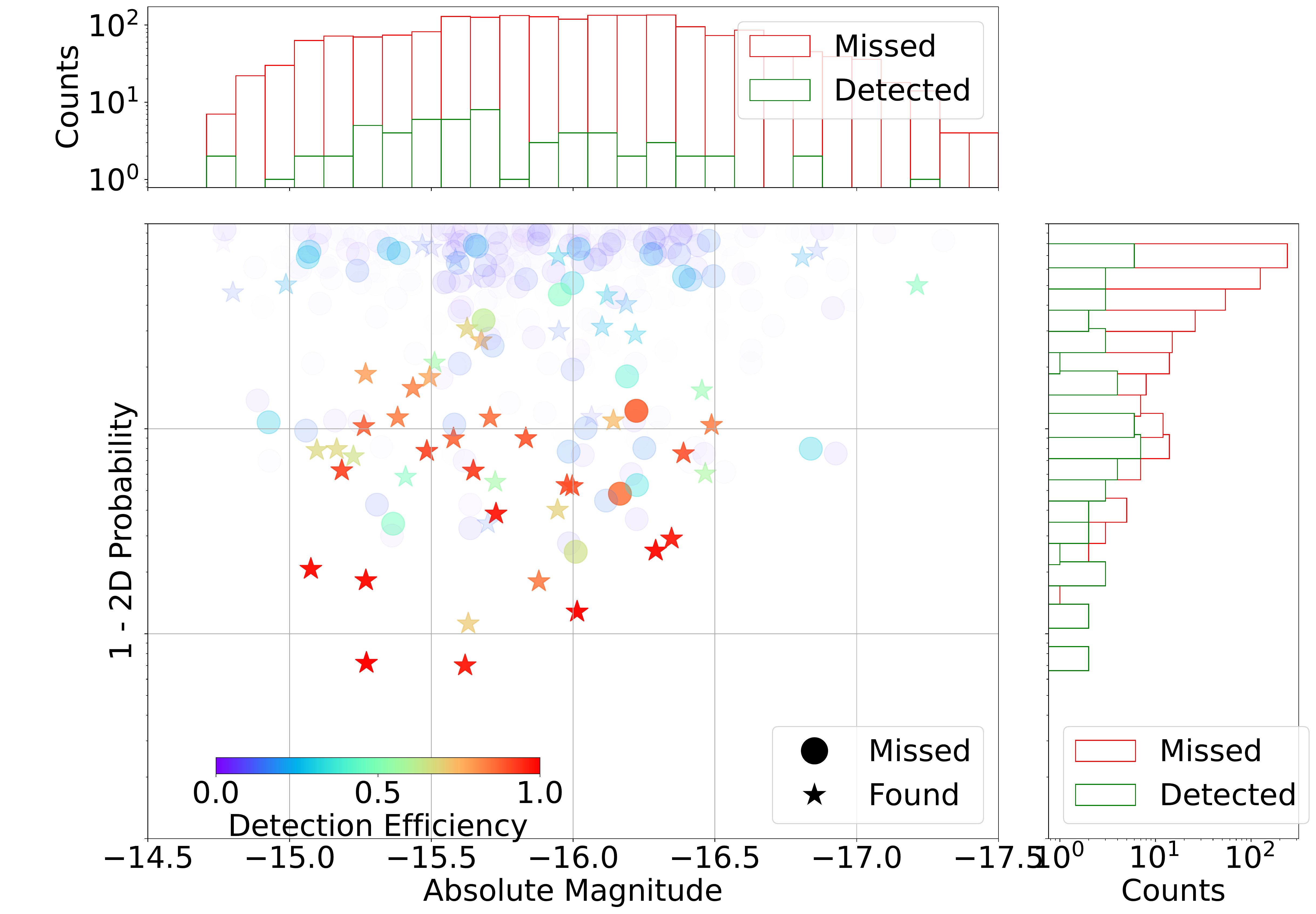}
    \includegraphics[width=3.5in]{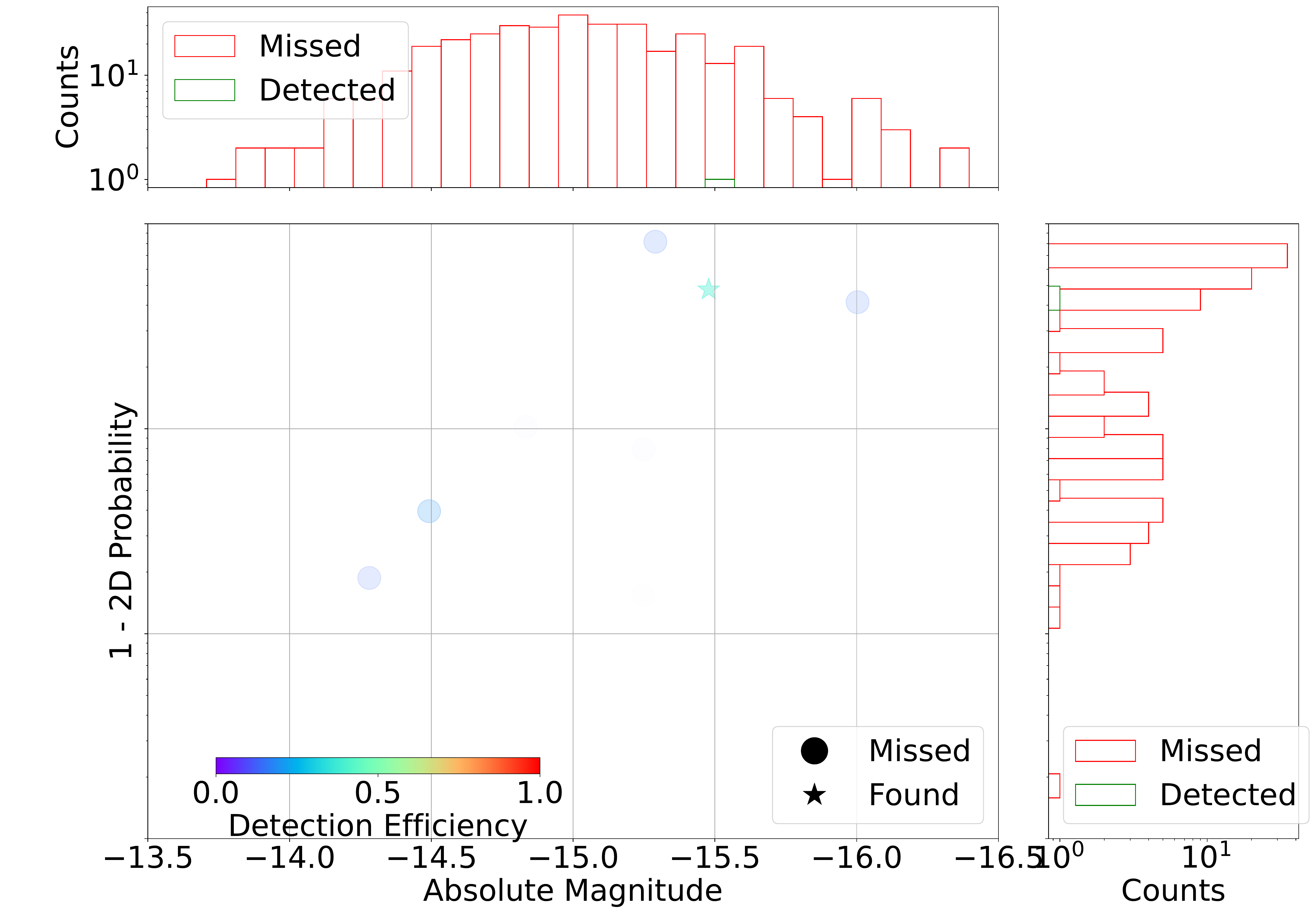}
    \includegraphics[width=3.5in]{figures/eff_NSBH_ZTF_O5.pdf}
    \caption{The plots at the top are BNS distributions of the missed and found injections(on the left BNS light curves observed by ZTF and the right one shows  the same observation in Rubin Observatory) , as a function of the peak $r$-band absolute magnitude and 2D-probability enclosed. We encode in the color bar and the transparency of the points the 3D-probability of finding the transient as measured by simulated injections. The one dimensional histograms are marginalized versions of the two dimensional scatter plot, with the detected set in green and missed set in red. The plots at  the bottom are the same for the NSBH light curves. Those plots concern the run O5}
    \label{fig:detection-O5}
\end{figure*}

In Figure \ref{fig:detection-O5}, we show the distribution of the missed and found Kilonovae (KNe) events for both the BNS and NSBH cases during observing run O5 with ZTF and Rubin Observatory observations, using Gravitational-Wave ElectroMagnetic OPTimization (\url{https://github.com/mcoughlin/gwemopt}).

\label{lastpage}
\end{document}